\documentclass[usenatbib]{mnras}  

\usepackage[T1]{fontenc}
\usepackage{ae,aecompl}

\usepackage{graphicx}
\usepackage{subfig}
\usepackage{amsmath}
\usepackage{amssymb}
\usepackage{pdflscape}
\usepackage{longtable}
\usepackage{supertabular}
\bibpunct{(}{)}{;}{a}{}{,} 
\usepackage{txfonts}
\def\Planck{\textit{Planck}}
\def\chandra{\textit{Chandra}}
\def\xmm{\textit{XMM-Newton}}
\def\HIFLUGCS{\textit{HIFLUGCS}}

 \title{Measuring the dynamical state of \Planck\ SZ-selected clusters:  \\
 X-ray peak - BCG offset }
\author[M.Rossetti et al. ]{M. Rossetti$^{1,2}$\thanks{E-mail: mariachiara.rossetti@unimi.it},
          F. Gastaldello$^{2}$,
          G. Ferioli$^{1}$, 
           M. Bersanelli$^{1}$,
           S. De Grandi$^{3}$,
           D. Eckert$^{4,2}$,
           \newauthor
           S. Ghizzardi$^{2}$,
           D. Maino$^{1}$
           \& S. Molendi$^{2}$\\
$^{1}${Dipartimento di Fisica, Universit\`a degli Studi di Milano,
  via Celoria 16, 20133, Milano} \\
$^{2}${INAF, Istituto di Astrofisica Spaziale e Fisica Cosmica, via Bassini 15,
20133 Milano}\\
$^3${INAF, Osservatorio Astronomico di Brera,  via E. Bianchi 46, 23807, Merate, Italy}\\
$^4${Astronomy Department, University of Geneva 16, ch. d'Ecogia, CH-1290 Versoix, Switzerland}
}

\begin{document} 
 \label{firstpage}
\pagerange{\pageref{firstpage}--\pageref{lastpage}}
\maketitle

\begin{abstract}
We want to characterize the dynamical state of galaxy clusters detected with the Sunyaev-Zeldovich (SZ) effect by \Planck\ and compare them with the dynamical state of clusters selected in X-rays survey.
We analyzed a representative subsample of  the \Planck\ SZ catalogue, containing the 132 clusters with the highest signal to noise ratio and characterize their dynamical state using as indicator the projected offset between the peak of the X-ray emission and the position of the Brightest cluster galaxy. 
We study the distribution of our indicator in our sample and compare it to its distribution in X-ray selected samples (HIFLUGCS, MACS and REXCESS). The distributions are significantly different and the fraction of relaxed objects is smaller in the \Planck\ sample ($52 \pm 4 \%$) than in X-ray samples ($\simeq 74\%$)
We interpret this result as an indication of different selection effects affecting X-rays (e.g. "cool core bias") and SZ surveys of galaxy clusters.        
\end{abstract}


\section{Introduction}
\label{sec:intro}
In the framework of hierarchical structure formation, clusters of galaxies, the largest and most massive collapsed objects in the Universe, represent the current endpoint of the evolution of primordial density fluctuations. Thus, they are at the same time sensitive probes of the history of structure assembly and powerful tools to constrain cosmological parameters. Indeed, much effort has been devoted in recent years to exploit the cluster population for cosmological studies, complementing other methods to break degeneracies between parameters.  However, the very same processes leading to the formation of clusters (i.e. accretion of smaller structures and mergers between objects with similar mass) may influence the results of cosmological studies, 
which often assume equilibrium and virialization, and should be properly taken into account \citep{cosmoPSZ1, cosmoPSZ2}. \\
Uncertainties in the scaling relations between observables at different wavelengths and the total mass of galaxy clusters have been shown to be the major source of systematics when using galaxy clusters as cosmological probes (e.g. \citealt{vikh09III, benson13, cosmoPSZ1, cosmoPSZ2}). 
Those uncertainties are at least partly associated to an incomplete knowledge of the physical processes affecting the baryonic components of galaxy clusters during their formation and evolution, which are not easily reproduced in cosmological simulations. Such processes are expected to play a role also in the selection of objects in clusters surveys: if they enhance (or decrease) the value of the observable used to find and select objects, the number of objects would be enhanced (or decreased) with respect to the expectation from the theoretical mass function. For instance, in X-ray surveys, the presence of a prominent surface brightness peak in the so-called "cool core'' clusters (i.e. clusters which are observationally defined by having a clear peak of X-ray emission associated to a decrease in the gas temperature, usually considered as relaxed objects) introduces a significant bias towards this class of objects \citep{eckert11}. \\
Growing attention has been devoted over the last decade to an alternative method to search for galaxy clusters: the Sunyaev-Zeldovich effect (SZ, hereafter, \citealt{sun70,sun72}), i.e. the distortion of the spectrum of the cosmic microwave background radiation induced by the Inverse Compton scattering of CMB photons on the electrons in the intracluster medium (ICM). The first large catalogues of galaxy clusters, containing hundreds of detections, have been published in recent years, using different instruments \citep{PSZ1, PSZ2, ACT_cat, SPT_cat}.  The main advantage of SZ surveys is that the SZ spectral distortion does not depend on the redshift of the source, allowing to construct virtually mass-limited samples and to eventually detect all massive clusters in the Universe, irrespective of their distance. Moreover, SZ quantities do not depend much on the details of the cluster physics and on the dynamical state of the cluster \citep{motl05, krause12, battaglia12}. Recently,  \citet{PSZ2} tested with Monte-Carlo simulations that the cluster morphology has a negligible impact on the source detection procedure in the \Planck\ survey. However, according to simulations \citep{pipino10,lin15}, the presence of a peaked pressure profile in cool core clusters results in an increase in the central value of the Comptonization parameter\footnote{We recall that the dimensionless Comptonization parameter, $y$, is proportional to the integral of the ICM pressure along the line of sight}, which could induce a bias in favor of cool cores also in SZ surveys. This effect is nonetheless expected to be small, especially for an instrument like \Planck\, whose spatial resolution is larger than the typical size of the cores of galaxy clusters and is thus more sensitive to the integrated total SZ signal rather than to its central value \citep{pipino10,lin15}.\\
On the observational side, only limited information is available yet on the properties of SZ-selected clusters, including their dynamical state. The majority of objects newly-discovered by \Planck\ show clear indication of morphological disturbances in their X-ray images, suggesting an active dynamical state \citep{planck_early_IX}, but a statistical analysis on the whole sample (or on a representative subsample) is necessary to draw any conclusion. 
For this reason, we performed the analysis described in the present paper which aims at measuring for the first time the dynamical state of a representative sample of \Planck\ SZ-selected clusters through an indicator of dynamical activity and compare it with the corresponding distribution for X-ray selected samples to answer to the following question: {\it is the cluster population selected through the SZ effect different, in terms of dynamical state, than the X-ray selected population?} \\
``Measuring'' the dynamical state of a cluster is not an easy task. In principle, the maximum amount of information on the dynamical history of a cluster can be derived by a detailed spatially-resolved two-dimensional mapping of thermodynamic quantities, of metal abundance distribution, associated to the study of the galaxy population, eventually to the presence of diffuse radio sources (halos and/or relics) and possibly to the mass distribution through gravitational lensing. However, this wealth of information is available only for a very limited number of objects and moreover it cannot be easily quantified in a single dynamical indicator.
The X-ray band alone can be successfully used to derive information on the dynamical activity, since merger events leave strong signatures in the thermodynamic quantities and morphological appearance of the ICM.  Powerful indicators assess the presence or absence of a cool core, such as central entropy \citep{cava09}, pseudo-entropy ratio \citep{leccardi10} or cooling time \citep{peres98}, but require spectroscopic analysis and eventually de-projection.  Less expensive indicators of dynamical activity can be computed basing only on the morphology of X-ray images, such as power ratios \citep{buotetsai}, centroid shifts \citep{poole06} and the concentration parameter \citep{santos08}.
An alternative approach to quantify the dynamical state can be built on the different physical processes undergone by the collisional ICM and the collisionless galaxy population during cluster mergers. 
In particular, brightest cluster galaxies (BCGs) are of particular interest as they represent a unique 
class of objects. They are the most massive and luminous galaxies in the Universe and their 
properties are found to correlate with many global cluster properties such as X-ray temperature
or luminosity \citep[e.g.][]{Edge:91,Edge.ea:91,Brough.ea:05,Brough.ea:08} indicating that their 
origin is closely related to that of the host cluster. 
If clusters are dinamically relaxed systems, we naturally expect the BCG to be at rest at the center
of the gravitational potential well, an assumption referred to as the ``central galaxy paradigm''
\citep[]{vandenBosch.ea:05}. However since the first X-ray images of clusters with the \emph{Einstein}
satellite became available, it became clear that there is a class of clusters for which BCGs are not close to the X-ray centres of their host clusters \citep{JonesC.ea:84,JonesC.ea:99}. The X-ray studies
complemented and supported the early evidence coming from the optical band that BCGs may not always 
be at the centre of the galaxy surface distribution \citep[e.g.][]{Beers.ea:83} and 
velocity space \citep[e.g.][]{Malumuth.ea:92,Oegerle.ea:01}. The connection between the presence
of offsets and the disturbed dynamical state of the cluster due to a merger has been progressively 
established in observational studies \citep{Katayama.ea:03,Patel.ea:06} and 
simulations \citep[e.g.][]{Skibba.ea:11*1}.
With the new generation of X-ray satellites, \emph{Chandra} and \emph{XMM-Newton}, it has become possible to strengthen the
correlation between the X-ray peak-BCG offset and a disturbed dynamical state (such
as lack of a cool core and disturbed X-ray morphology 
\citealt{sanderson09,hudson10,mann_ebe,Hashimoto.ea:14}) and to establish this indicator as a simple but robust diagnostic of an active dynamical state. Sometimes the different flavor
of using the X-ray centroid rather than peak is used, but leading to basically the same results 
\citep{mann_ebe}. \\
In this paper, we measure the offset between the X-ray peak and the BCG population as indicator of dynamical state of \Planck\ SZ-selected clusters  and compare its distribution to the one of X-ray selected samples  to provide a first answer to the question we posed above. Therefore we can re-formulate the aforementioned question as: {\it is the distribution of the BCG-X ray peak offset different in the \Planck\ SZ survey different than in X-ray selected samples?} This question is obviously less ambitious than our starting question but it represents a first significant step towards a more complete characterization of the population of clusters selected through the SZ effect. \\
In this paper we assume $\Lambda$-CDM cosmology with $H_0=70 \rm{km}\,\rm{s}^{-1}\, \rm{Mpc}^{-1}$, $\Omega_m=0.3$ and $\Omega_\Lambda=0.7$. The outline of the paper is as follows: in Sec. \ref{sec:analysis} we introduce our sample and describe the procedure we used to measure our indicator. In Sec. \ref{sec:distr}, we describe its distribution in the \Planck\ sample and compare it to X-ray selected samples in Sec. \ref{sec:xsamples}. We discuss our findings and provide an interpretation in Sec. \ref{sec:discussion}.
\section{Data analysis}
\label{sec:analysis}
\subsection{The sample}
The starting point of our analysis is the \Planck\ cosmology sample (PSZ1-cosmo) described in \citet{cosmoPSZ1}.
It is a high-purity subsample constructed from the first release of the \Planck\ catalogue of SZ sources \citep{PSZ1}, by imposing a signal-to-noise ratio ($S/N$) threshold of 7 and applying a mask, that excludes the galactic plane and point sources leaving $65\%$ of the sky for the survey. It contains 189 bona-fide clusters with associated redshifts and has been used for the cosmological analysis with cluster number counts described in \citet{cosmoPSZ1}. The first release of the Planck SZ catalogue (PSZ1, hereafter) has benefited from a massive multi wavelength follow up campaign to confirm the detected candidates, measure their  redshifts and characterize the sample.  More specifically the cosmology sample has been almost completely followed-up in X-rays with either \chandra\ or \xmm\, allowing us to have a reliable estimate of the peak position (see Sec. \ref{sec:xray}).
 Since a similar campaign has not been possible yet for the larger and more recent second release of the Planck SZ catalogue (PSZ2, \citealt{PSZ2}), we decided to base our analysis on the PSZ1 catalogue. \\
Unfortunately, we do not have literature information concerning the BCGs of all the clusters in the PSZ1 cosmological sample (Sec. \ref{sec:BCG}) and not all the X-ray observations are public yet. In order to minimize the number of clusters lacking the offset measurement, we decided to extract a subsample from the PSZ1-cosmo, by imposing $S/N>8$. We decided to cut in signal to noise to reproduce as closely as possible the selection function of \Planck\ SZ surveys. With such more stringent $S/N$ threshold, our final sample is composed of 136 objects: except for four objects lacking X-ray observations (Sec. \ref{sec:xray}), we could measure the BCG-peak offset for the remaining 132 clusters. We verified that our sample is representative of the parent PSZ1-cosmo sample by performing a Kolmogorov-Smirnov test on their distributions of redshifts (probability that they are drawn from the same parent distribution $p_0=0.97$) and masses ($p_0=0.82$). \\
We provide the list of clusters in our sample in Table  \ref{tab:maintable}, where we list the index and name in the PSZ1 catalogue, the redshift and the angular size $\Theta_{500}$, corresponding to $R_{500}$.
We estimated this latter quantity using the redshift and masses in the updated PSZ1 catalogue \citep{PSZ1_update}, which were obtained with the $Y-M$ scaling relation in \citet{cosmoPSZ1}. \\

\subsection{Determining the X-ray peak}
\label{sec:xray}
We determined the coordinates of the X-ray peak using X-ray images obtained with the last generation high-spatial resolution X-ray telescopes, preferentially \chandra.
We downloaded the \chandra\ images of 125 clusters, identified bright point sources, smoothed the images with a Gaussian function with $FWHM=3-5$ pixels and mark the position of the brightest pixels. Seven clusters in our sample were not observed with \chandra\ but had public {\it XMM-Newton} observations, that we used to estimate the peak position. 
We could not determine the position of the X-ray peak for four clusters in our reduced \Planck\ sample of 136 objects which have been observed by \chandra\ but whose observations are not public yet.  The absence of this very small number of clusters from our sample does not introduce any foreseeable bias, as these four objects are not peculiar in terms of redshift and mass and they are not new \Planck\ discovered objects.\\
In principle, the superb angular resolution of \chandra\ allows us to estimate the position of the peak of the X-ray emission with great accuracy, $<0.3$ arcsec \citep{evans10}, which at the median redshift of our sample corresponds to  $< 8$ kpc. However, as discussed in \citet{mann_ebe}, the accuracy on the position also depends on the statistical quality of the observations, on the possible presence of non-detected point sources and on the surface brightness distribution (i.e. presence of multiple peaks). It is thus not easy to estimate this uncertainty for all clusters and the astrometric error reported above should be considered only a lower limit. Moreover in 7 cases we could not use \chandra\ observations but used the lower resolution \xmm\ data which are characterized by a larger positional error: we verified a posteriori that the use of these instruments does not affect our conclusions by excluding them from our sample and finding consistent results.  However, the uncertainty in the positional reconstruction is not a systematic error, as it will not produce systematically larger or smaller offsets. Indeed, in the few cases where two possible peaks were detected (as for instance in double systems or in the presence of infalling subclusters) we always chose the brightest pixel, regardless of its proximity to the BCG. \\
\subsection{Finding the BCG}
\label{sec:BCG}
We based our search for BCGs mainly on literature information: optical catalogues of galaxy clusters which provide the position of the BCG (MaxBCG \citealt{maxbcg}, Wen12 \citealt{wen12}, redMaPPer\footnote{In the redMaPPer algorithm the centring of the clusters is fully probabilistic, to take into account multiple candidate central galaxies. We thus verified one by one the associations with redMaPPer clusters.} \citealt{rykoff14}) and papers providing the position of the BCGs for different samples of galaxy clusters (\citealt{coziol09}, \citealt{hoffer12} for ACCEPT, \citealt{crawford99} for ROSAT-BCS, \citealt{zhang11} for HIFLUGCS, \citealt{mann_ebe} for MACS\footnote{Although the BCG coordinates are not provided in the paper by \citet{mann_ebe}, they were kindly provided us by H. Ebeling (private communication).}, \citealt{song12} for SPT and \citealt{menanteau10} for ACT clusters). We first cross matched our sample with the BCG catalogues listed above using TOPCAT \citep{topcat} and associated a BCG to 98 clusters. We evaluated case by case the objects where two different BCGs were associated by different catalogues to the same cluster (the most relevant examples are provided in Appendix A) and we selected as BCG the brightest one in the NED \footnote{https://ned.ipac.caltech.edu/} database.\\
For 38 objects we could not find any information in the catalogues and papers listed above.  We thus searched in NED for galaxies around the X-ray position in a circle with radius $R_{500}$. In 18 cases, one of the galaxies (the brighest in the list) was cited as BCG in one or more literature works (optical studies of individual objects or BCG catalogues for smaller sample of clusters). We associated those BCGs to their clusters and refer to the papers who made that association in Table \ref{tab:maintable}. \\
 For the remaining 16 clusters, which are all out of the sky region covered by the Sloan Survey, we made our own choice of the BCG as the brightest source (using 2MASS magnitudes) classified as galaxy in the objects found by the NED database within $R_{500}$ of each cluster. We then visually inspected the Digitized Sky Survey (DSS) images of those clusters to confirm the identification of the BCGs .\\
We could associate a BCG to all clusters in our sample and therefore measure the projected offset between the  BCG and the X ray peak ($D_{X-BCG}$ hereafter) for 132 clusters. In  Table \ref{tab:maintable} we provide the coordinates of the BCG and the X-ray peak as well as our measured $D_{X-BCG}$ in arcsec, kpc and fractions of $R_{500}$. \\
The optical information from which we derived the positions of the BCGs are very heterogeneous and it is thus difficult to estimate the uncertainties in our measurements. First of all, different data-sets have different absolute astrometric accuracy. Secondly, different choices and methods (optical selection, searching radius, colors) made by the Authors of the references we used, introduce an uncertainty in our data which is likely dominant over the error on the galaxy position. In a few cases, some of the literature work we have used for BCG association may have induced a systematic bias in our analysis by limiting the BCG search in a radius smaller than $R_{500}$ (e.g. \citealt{hoffer12} search the BCG in a $5^\prime\times5^\prime$ field-of-view centered on the X-ray position, which is often smaller than a circle with radius $R_{500}$ for low-redshift systems) or by choosing the BCG closer to the X-ray peak in systems with two or more galaxies with comparable magnitudes.  Therefore, it is possible that in a few cases our offsets may be underestimated. 

\section{Results}
\label{sec:results}
\subsection{The offset distribution of the \Planck\ sample}
\label{sec:distr}
\begin{figure*}
\centering
\includegraphics[width=0.45\textwidth]{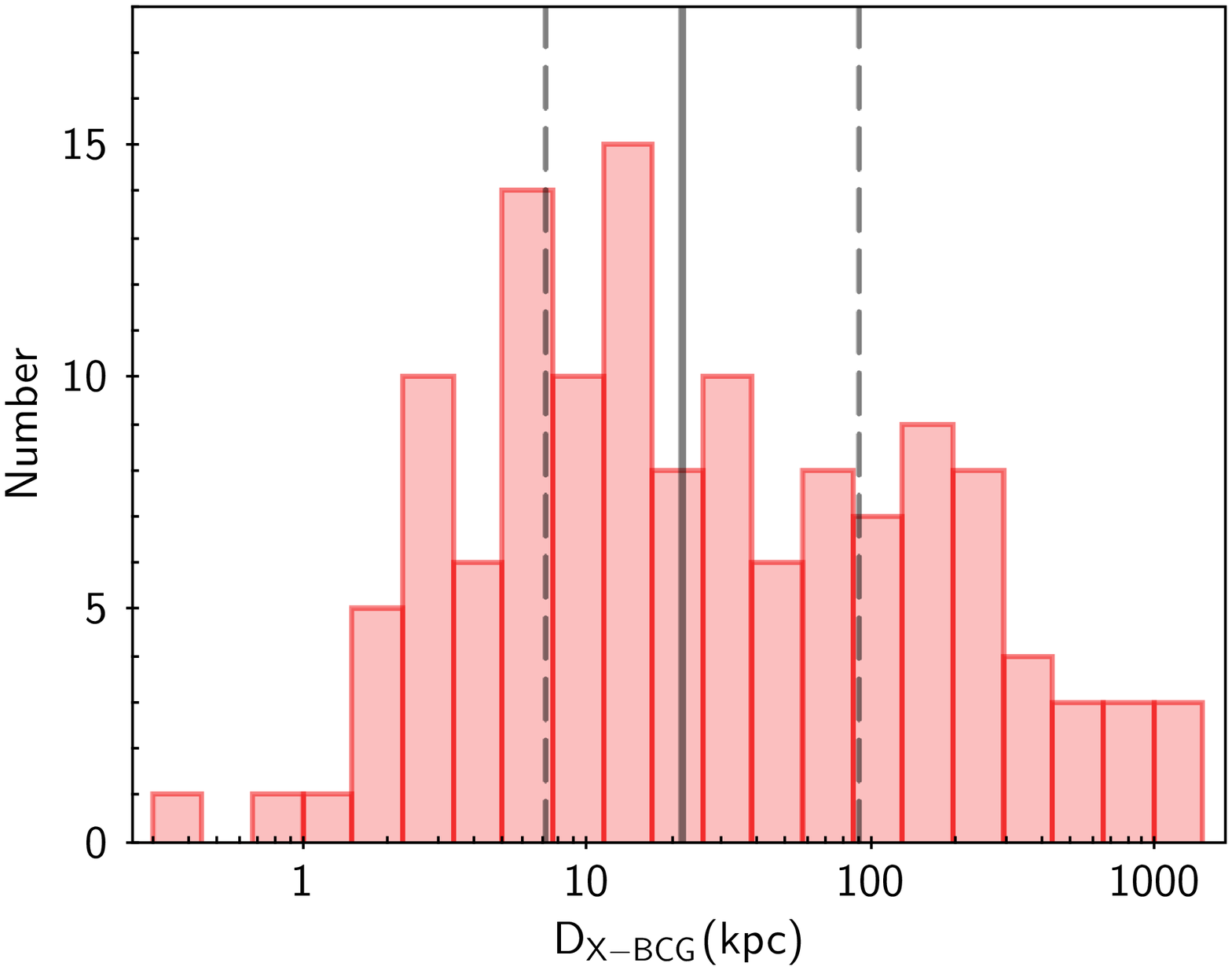}
\includegraphics[width=0.45\textwidth]{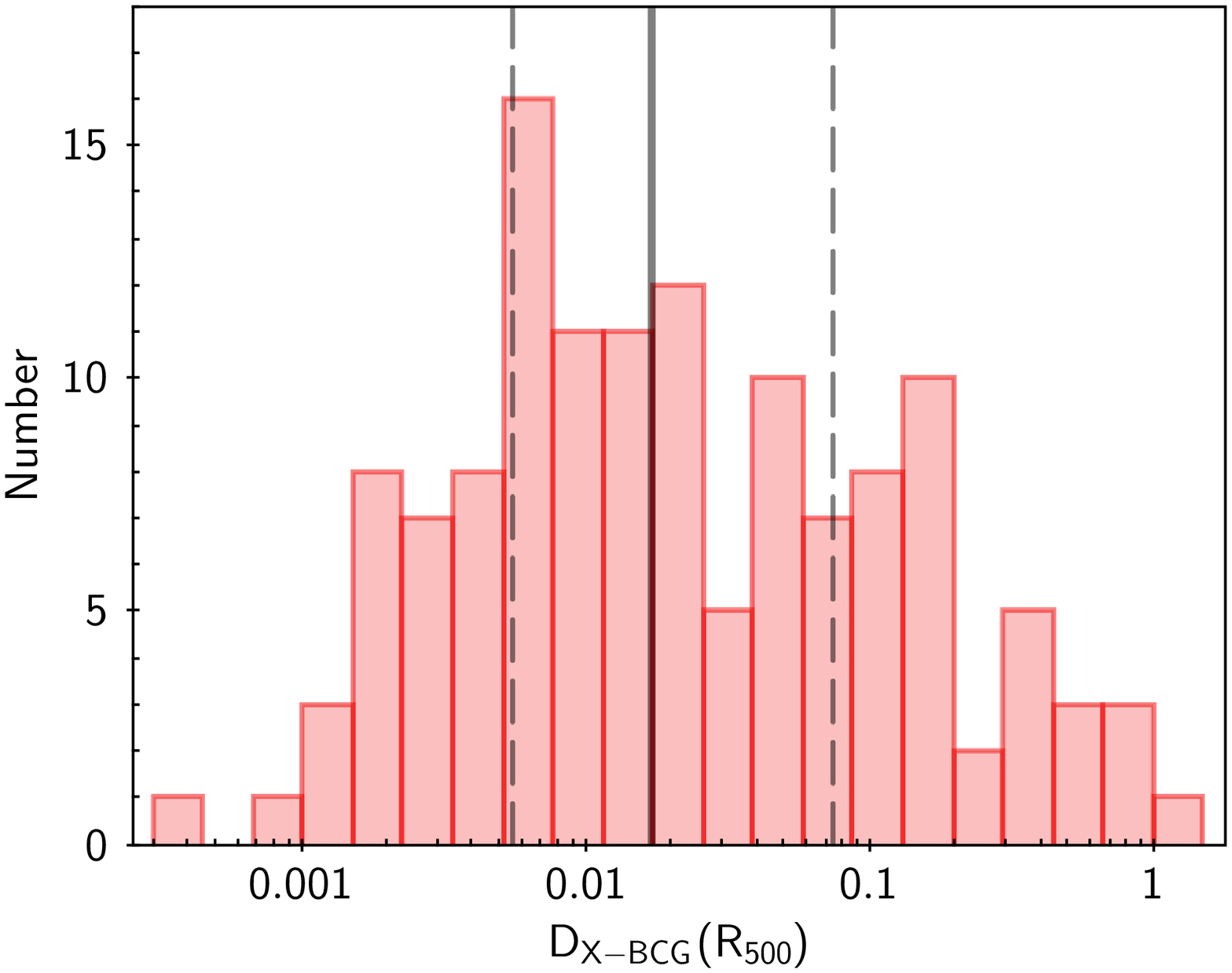}
\caption{Offset distribution between the X-ray peak and the BCG
  position in units of  kpc (left) and $R_{500}$  (right) for our \Planck\ sample. The grey line indicates the median of the distributions and the dashed lines the first and the third quartiles. }
  \label{fig:histo}
\end{figure*}

As described in Sec.\ref{sec:analysis}, we could measure the offset between the peak of the X-ray emission and the BCG position for  our sample of 132 \Planck\ selected clusters. In Fig.\ref{fig:histo}, we show the distribution of our indicator $D_{X-BCG}$ both in units of kpc and rescaled by $R_{500}$.   The shape of the distribution is roughly lognormal, with a median value $0.017 R_{500}$ ($21.5$ kpc) and has a large spread which can be expressed in terms of the Interquartile Range\footnote{The interquartile range is an indicator of the statistical dispersion of a distribution and is defined as the difference between the third (75th percentile) and the first (25th percentile) quartiles.} (IQR$=0.066 R_{500}$ or $82.4$ kpc).  When plotted in logarithmic scale, the distribution is not symmetric around the maximum but skewed towards large offset values. Indeed, a significant number of objects feature separations of the order of hundreds of kpc and of large fractions of $R_{500}$. \\
The distribution in Fig. \ref{fig:histo} is not bimodal and does not provide us a clear threshold to divide clusters in two separate classes, ``relaxed''  and ``disturbed''. In the literature, \citet{mann_ebe} classify   
objects with an offset $>42$ kpc as ``extreme mergers'' and 50 clusters in our sample ($38\%$ of the total) would fall in this class. However, given the relatively large mass range in our sample  (covering about an order of magnitude) and since we want to compare it with other samples (Sec. \ref{sec:xsamples}), we prefer to define a more physically interesting threshold in terms of $R_{500}$. \citet{sanderson09} divide their objects into two classes:  ``small offset'' ($<0.02R_{500}$) systems, which can be considered as relaxed, and  ``large offset'' ($>0.02R_{500}$) systems which are likely disturbed. We decided to follow this convention and in the rest of this paper we define as `relaxed'' the 68 objects where the offset is smaller than $0.02R_{500}$. We thus find a fraction of relaxed object in our sample of $(52\pm 4)\%$, where we estimated the error with bootstrap resampling.\\
We divided our sample into two halves, first a ``low-redshift'' and a ``high-redshift'' subsamples (splitting around the median value $z=0.16$) then a ``low-mass'' and a ``high-mass'' subsamples (around the median value $M_{500}=6.4\times 10^{14}M_\odot$). We compare the offset distributions in units of $R_{500}$ for the different subsamples. We calculated for each subsample the fraction of relaxed objects and found $64\%$ for the low$-z$ and $39\%$ for the  high$-z$, $62\%$ for the low-mass and $41\%$ for the high mass, with an uncertainty of $6\%$ in each subsample. The difference in the relaxed fraction between low$-z$ and high$-z$ is significant at $2.8\sigma$ and provides some indication of an evolution with redshift. We observe a slightly less significant, but still tantalizing, difference ($2.5\sigma$)  between the low mass and high-mass subsamples. 
We may further compare the subsamples by trying to asses the probability that they are drawn from the same parent distribution. This is a classical problem in statistics and since we do not know the underlying distribution we resort to non-parametric test. We  follow the advice of \citet{wall2003} (see their table 5.6) and choose the most efficient 
non-parametric tests: the Kolmogorov-Smirnov (KS) two-sample test and
the Wilcoxon-Mann-Withney (WMM) $U-$test. The KS test in its two-tailed version 
applied in this study is sensitive to any form of difference between the 
two distributions. The $U-$test is sensitive to the position of the 
distributions, i.e. location of means and medians.  We follow the suggestion of \citet{feigelson2012} and compare the results of more than one method, as various tests have different efficiencies under various conditions. We apply the above tests using the R environment for statistical computing \citep{r_cite} and we show our results are in the upper part of Table \ref{table:ks}: we find significant indication (null hypothesis probability $p_0<0.2\%$) that the distribution is different in the two redshift subsamples and some indication ($p_0<4\%$) in the two mass bins. Given the limited number of objects in our sample, especially at high redshifts and mass, we cannot divide our sample in more mass and redshift bins, otherwise we would be dominated by statistical uncertainty. Moreover, there is a significant overlap between our low-redshift and low-mass subsamples, as well as in the high-mass and high-$z$, because the least massive objects are detected only locally in the \Planck\ survey (see the distribution of objects in the mass-redshift plane in the PSZ1 \citealt{PSZ1}). Therefore it is not possible to assess if we are observing a dependence of the relaxed fraction on the mass, redshift or both.
\setcounter{table}{1}
\begin{table}
 \centering
\begin{tabular}{l c c c c }
\hline
Compared samples & \multicolumn{2}{c}{KS test} & \multicolumn{2}{c}{MWW test} \\
& $D$ & $p_{0}$ & $U$ & $p_{0}$ \\
\hline
\Planck\ redshift bins &  $0.323$ & $2.1\ 10^{-3}$ & $2823$ &  $9.5\ 10^{-4}$ \\
\Planck\ mass bins & $0.246$ & $3.8\ 10^{-2}$ & $2555$ & $3.9\ 10^{-2}$ \\
\hline
\Planck--HIFLUGCS & $0.336$ &  $1.1\ 10^{-4}$  &  $5440$ & $5.6\ 10^{-5}$   \\
\Planck--MACS & $ 0.228$ &  $4.2\ 10^{-3}$ & $8865$ & $4.4\ 10^{-4}$ \\
\Planck--REXCESS & $0.297$ & $2.2\ 10^{-2}$ & $2637$ & $3.9\ 10^{-2}$ \\
\hline
\Planck--MACS high$-z$ & $0.375$ & $1.6\  10^{-5}$ & $4903$ & $8.7\ 10^{-7}$ \\
\Planck--MACS high$-M$ & $0.336$ & $1.2\ 10^{-3}$ & $2720$ & $1.4\ 10^{-4}$ \\
\hline
\end{tabular}
\caption{Results of statistical tests comparing two distributions. The first two lines refer to the comparison between redshift and mass subsamples of the \Planck\ sample (Sec. \ref{sec:distr}). The middle lines refer to the comparison between our \Planck\ samples and the X-ray selected samples (Sec. \ref{sec:xsamples}). The bottom lines compare only the high-redshift and high-mass subsamples of \Planck\ and MACS to assess the origin of our results (Sec. \ref{sec:sz_vs_x}). }
\label{table:ks}
\end{table}
\subsection{Comparison with X-ray selected samples}
\label{sec:xsamples}
\begin{figure*}
\centering
\hspace{-0.5cm}
\includegraphics[width=0.5\textwidth]{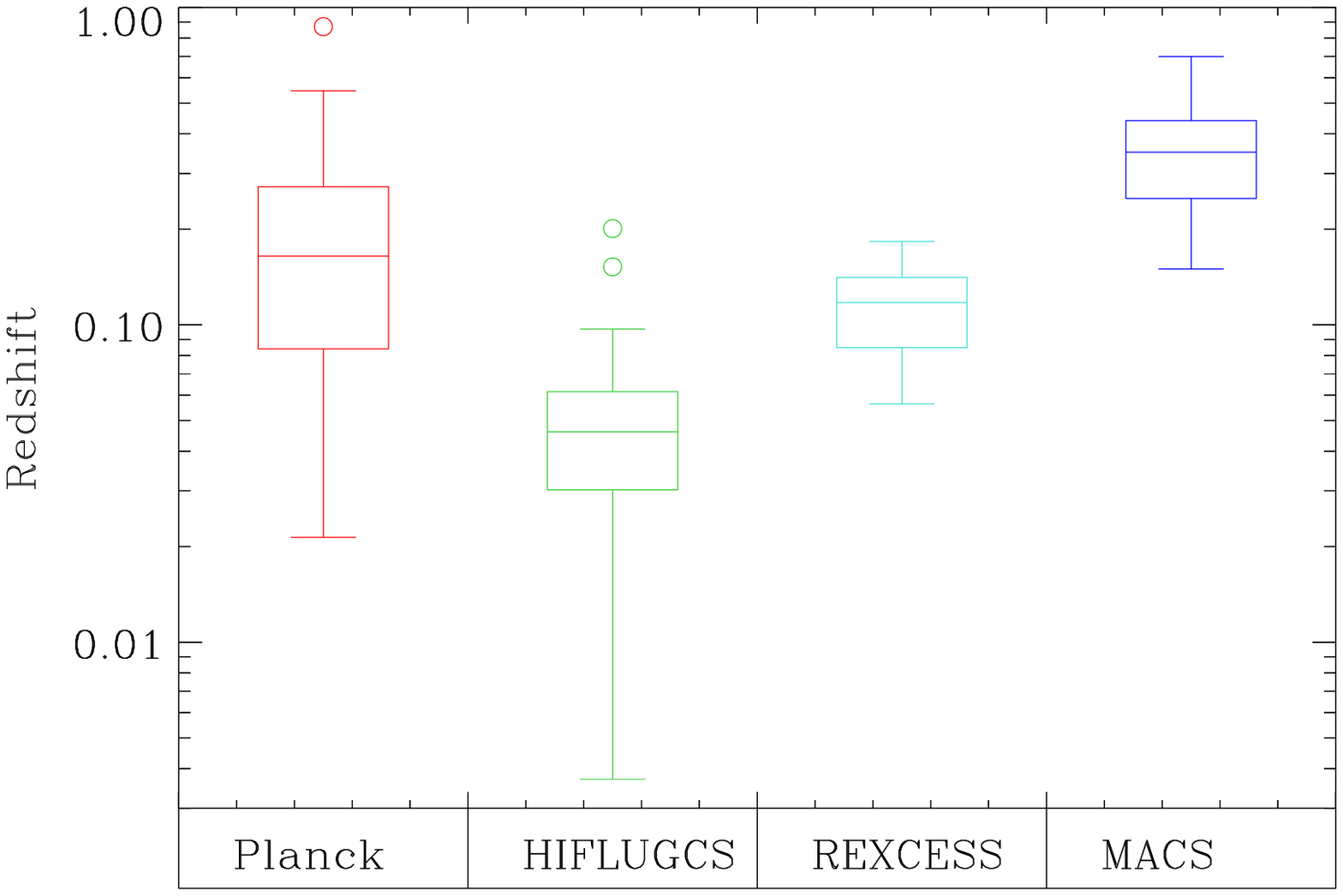}
\includegraphics[width=0.5\textwidth]{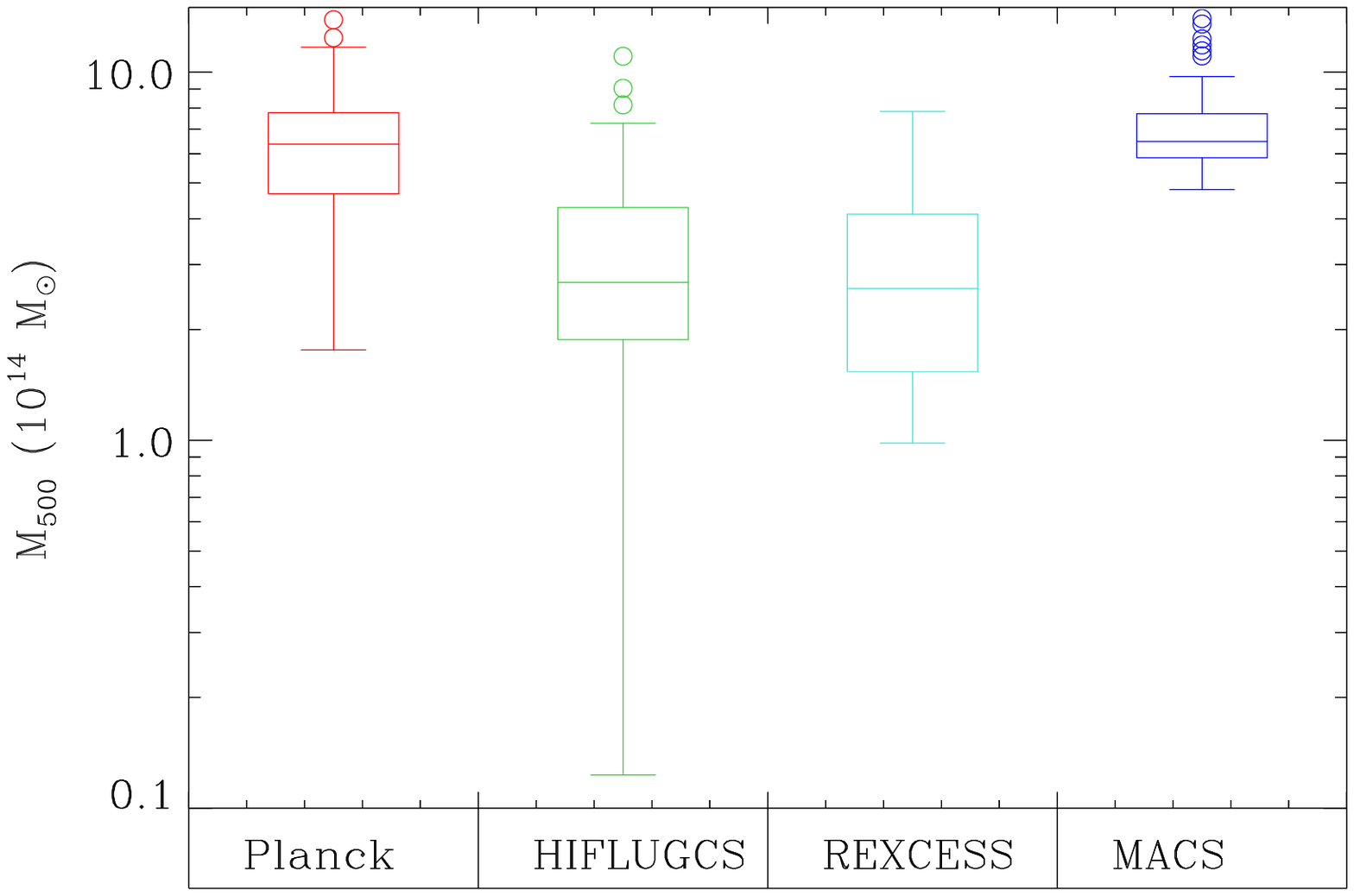}
\caption{Box-and-Whiskers plot representing the redshift (left) and mass (right) distribution of the \Planck\ sample compared to the  three X-ray selected samples. The empty circles mark outliers in the distribution, selected for being values larger than the 75th percentile $+1.5IQR$.
The properties of the Planck sample are intermediate between the low-redshift low-mass
  objects of the \HIFLUGCS\  and REXCESS sample and the high mass high-$z$
  distribution of MACS.}
  \label{fig:boxplot}
\end{figure*}
\begin{figure*}
\subfloat[][]{\includegraphics[width=0.3\textwidth]{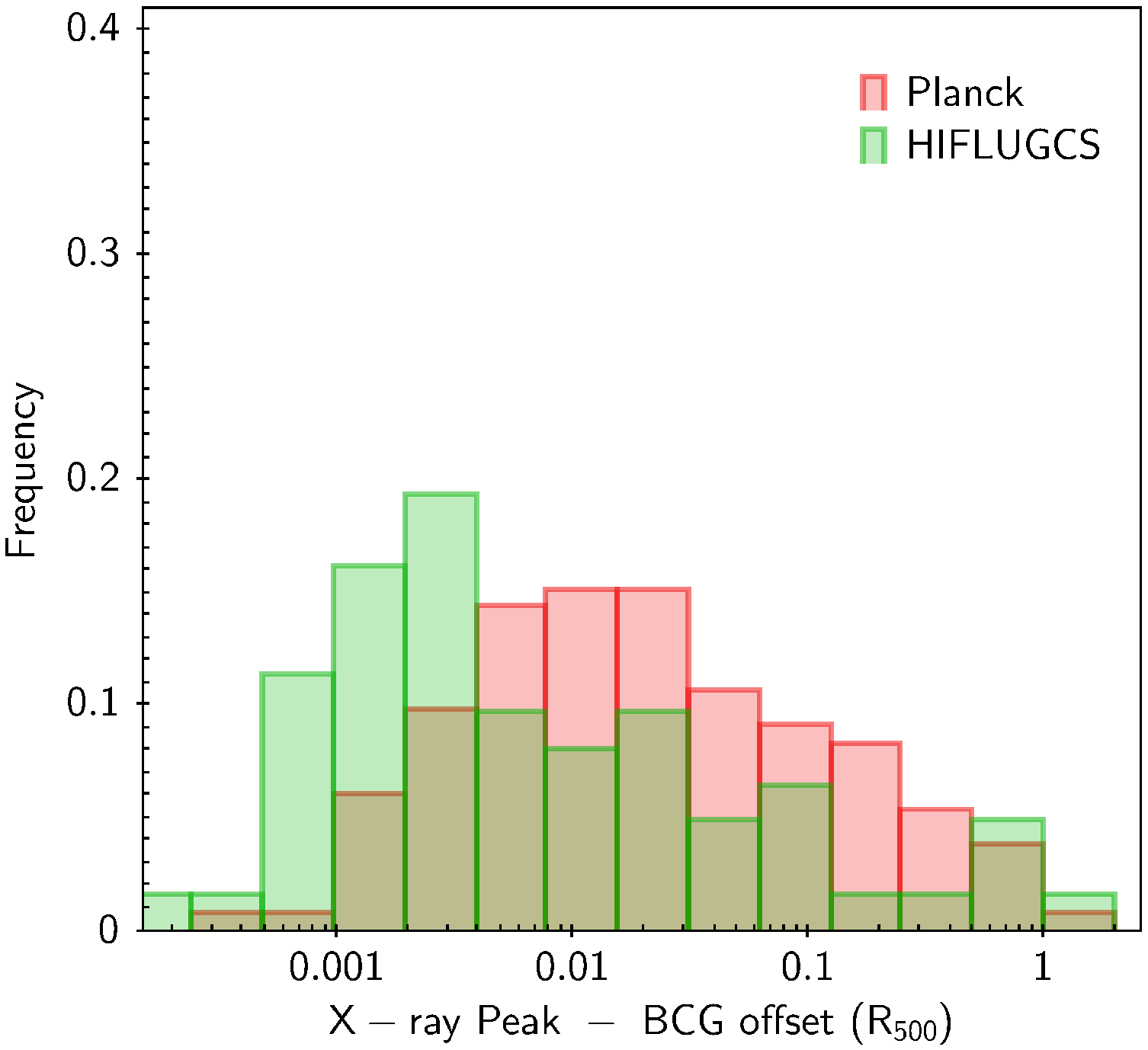}}
\qquad
\subfloat[][]{\includegraphics[width=0.3\textwidth]{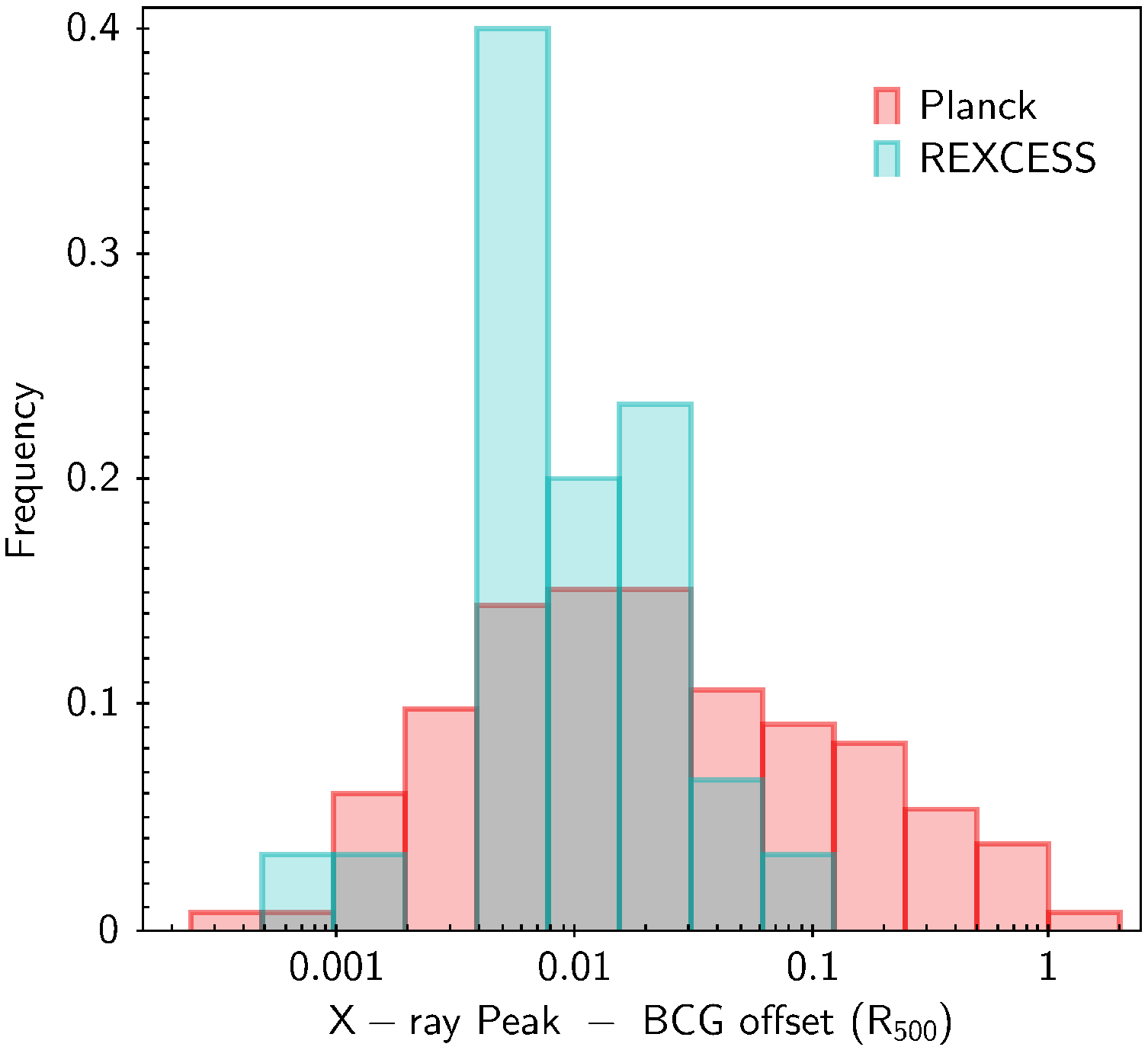}}
\qquad
\subfloat[][]{\includegraphics[width=0.3\textwidth]{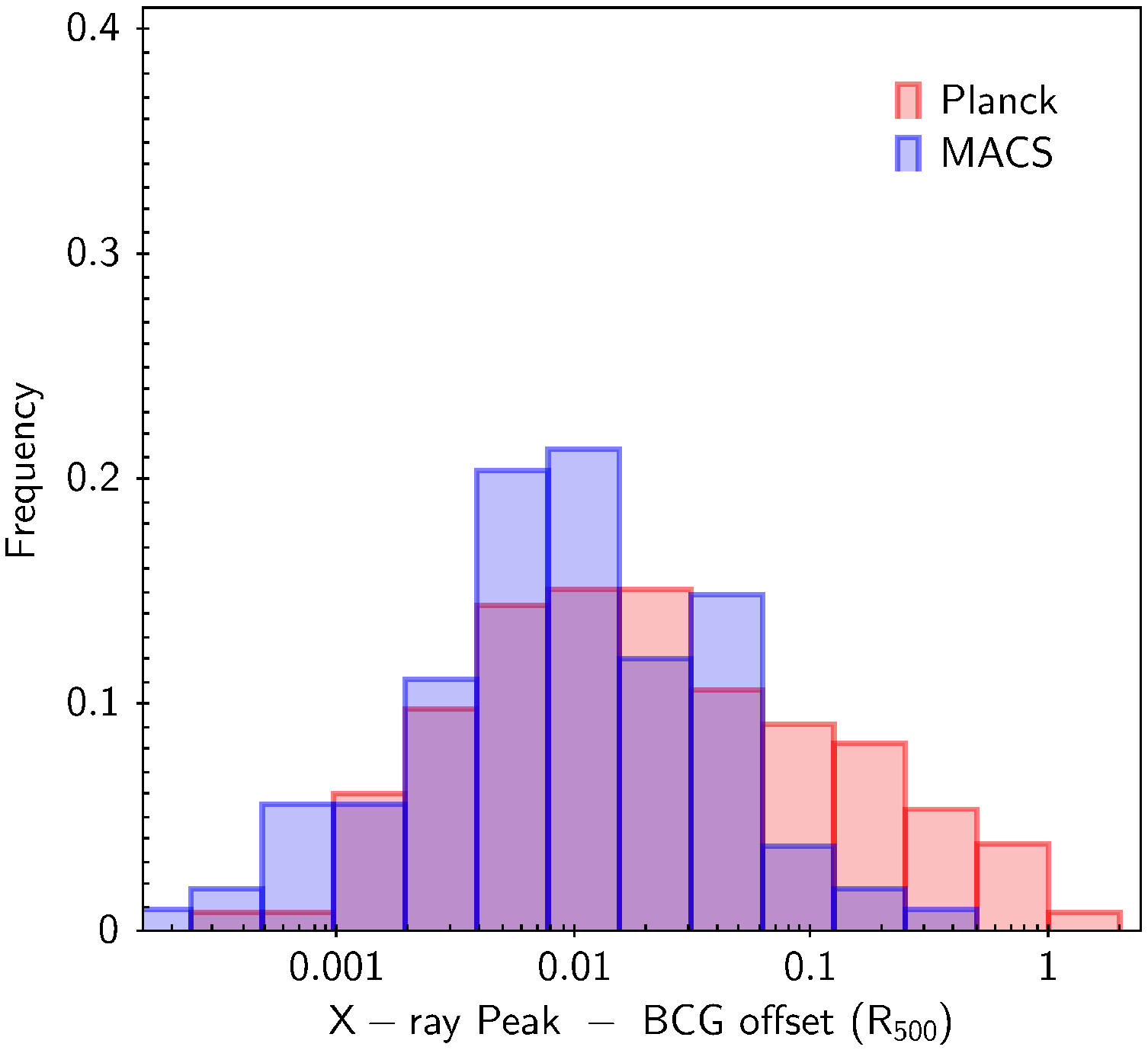}}\\
\subfloat[][]{\includegraphics[width=0.3\textwidth]{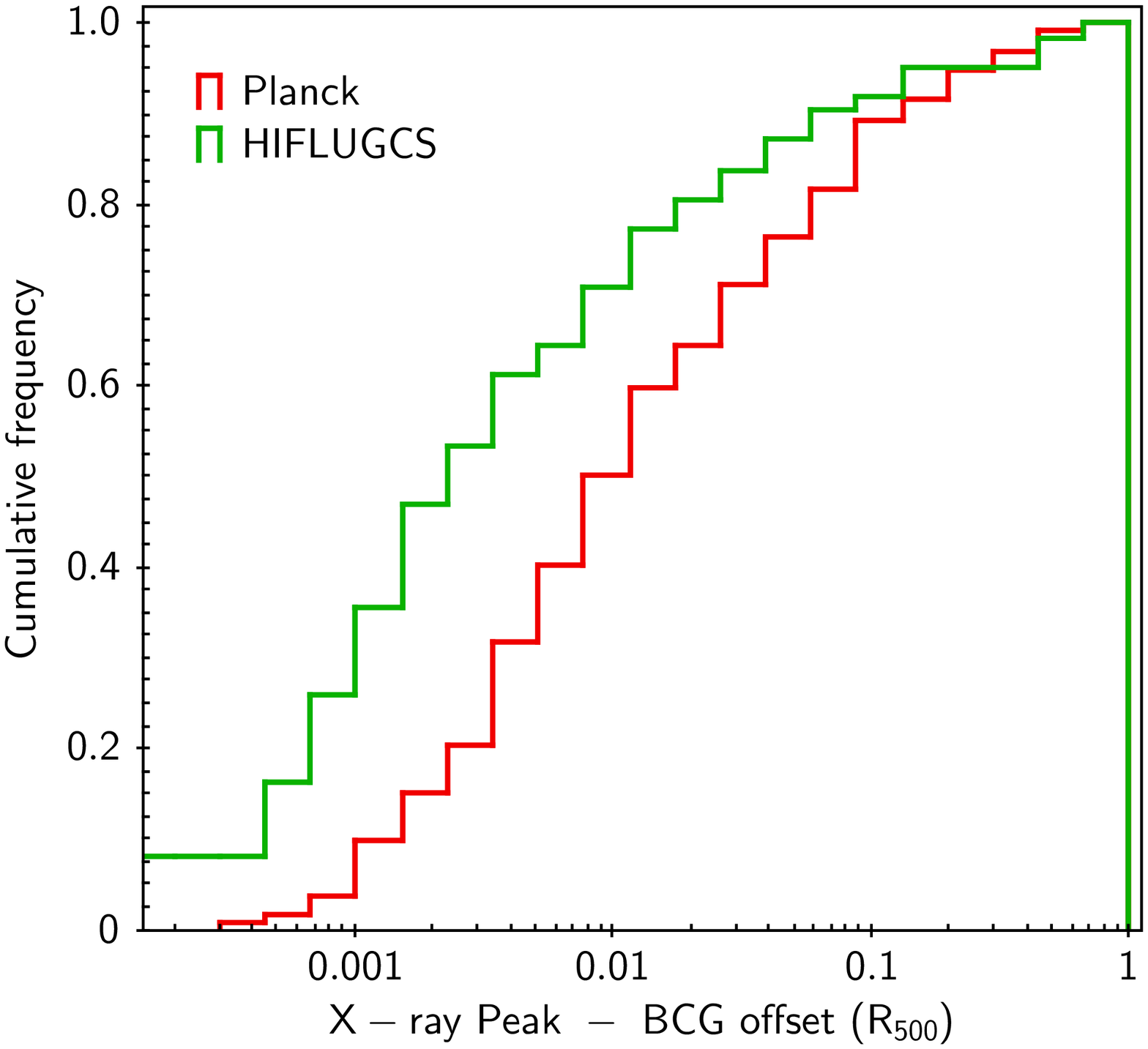}}
\qquad
\subfloat[][]{\includegraphics[width=0.3\textwidth]{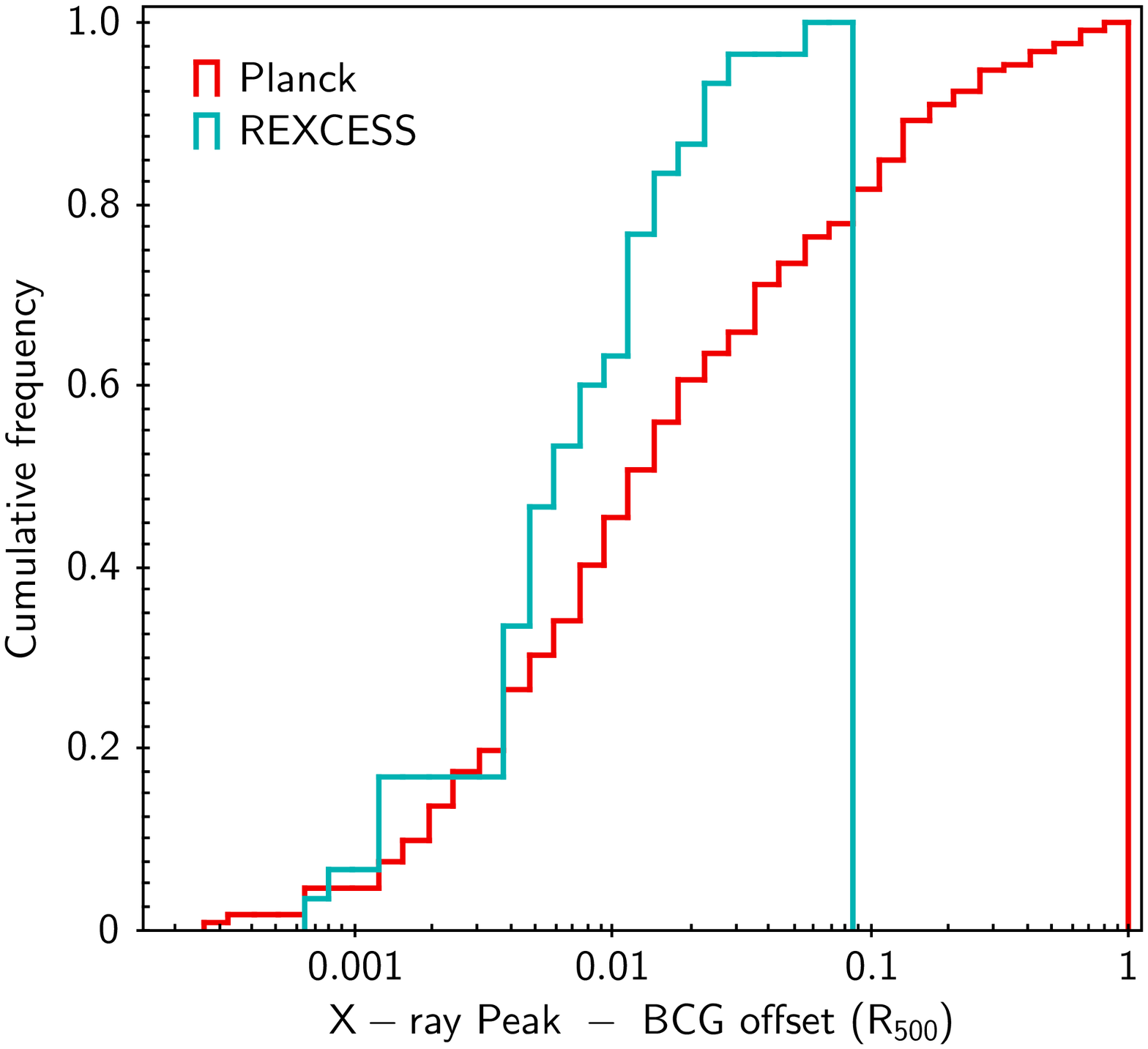}}
\qquad
\subfloat[][]{\includegraphics[width=0.3\textwidth]{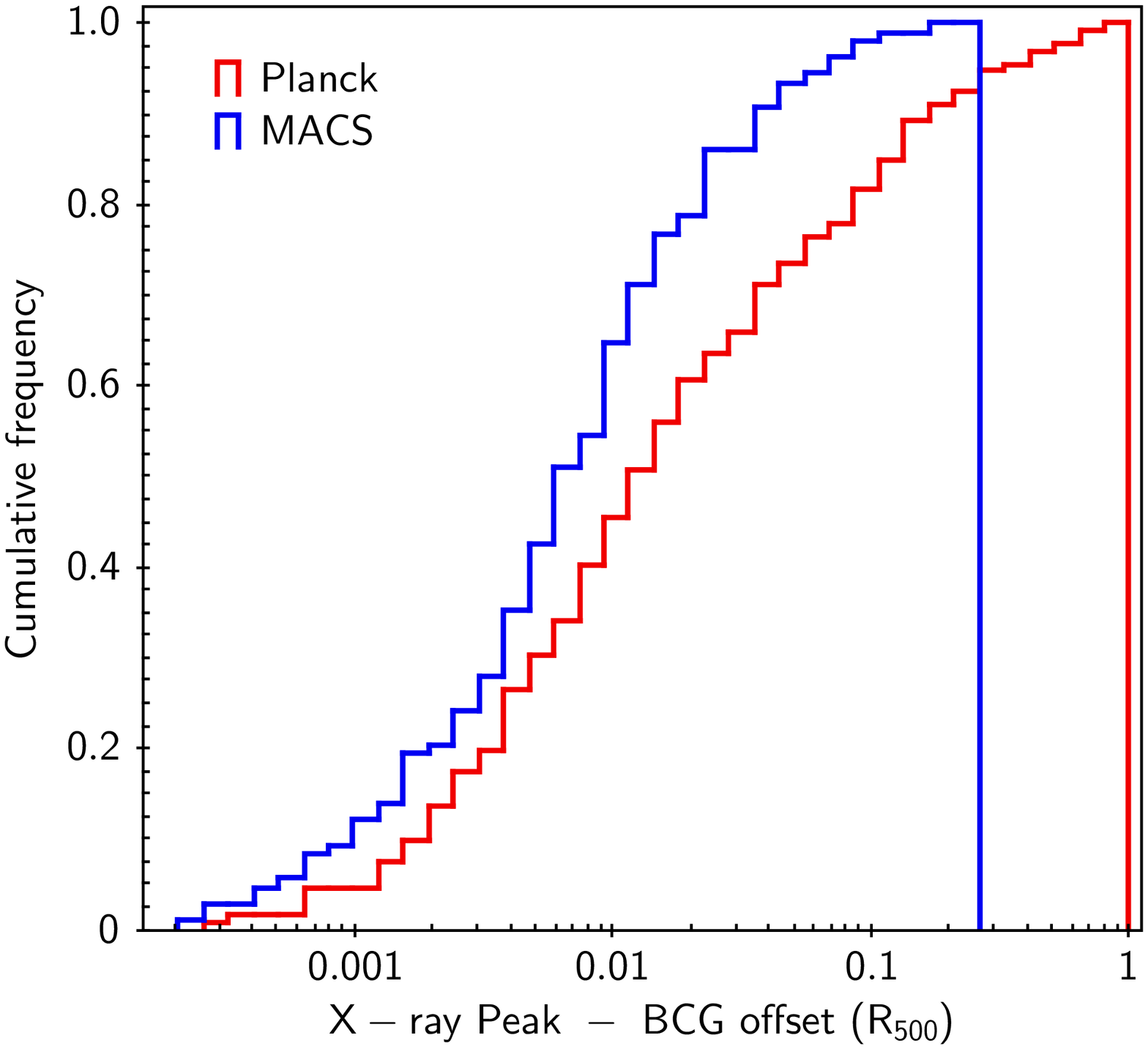}}
\caption{Distribution of the offset between the X-ray peak and the BCG for the \Planck\ sample (red) compared with \HIFLUGCS\ (green, left), REXCESS (cyan, middle) and MACS (blue, right). In the top row we show the normalized histograms and in the bottom row the cumulative distribution.  }
\label{fig:xray_comp}
\end{figure*}

In order to answer the question we asked in Sec. \ref{sec:intro}, we need to compare the offset distribution that we obtained for our sample with a consistent distribution for X-ray selected samples. The offset distribution has been studied in the literature by many authors for several samples of galaxy clusters \citep{lin_mohr, sanderson09, zhang11, mann_ebe, haarsma10, stott12}. However, since we want to compare the offset distribution of SZ selected clusters with X-ray selected clusters, we compare our distribution only with the  samples that ensure a rigorous selection criterium based on X-ray surveys. Moreover, as shown in \ref{sec:distr}, the $D_{X,BCG}$ distribution may evolve with redshift and mass, so ideally we would like to compare our \Planck\ sample with an X-ray selected sample with the same redshift and mass distribution. However, such a sample does not exist because of the different selection functions in the mass-redshift plane of SZ and X-ray surveys. Therefore we decided to compare our sample with three X-ray selected samples, with different mass and redshift ranges (Fig. \ref{fig:boxplot}), which are described below.
\begin{itemize}
\item {\it HIFLUGCS} \citep{reiprich02} is a complete flux-limited sample, comprising the 64 X-ray brightest clusters  ($F_X[0.1-2.4\ \rm{keV})]> 2 \ 10^{-11}\ \rm{erg\ s}^{-1}\rm{cm}^{-2}$) outside of the galactic plane.
The position of their BCGs is reported in  \citet{zhang11}, who base their analysis on optical data obtained within an aperture $>2.5$ Mpc, larger than the typical $R_{500}$ of their clusters.
As  \citet{zhang11} measure their offset from the X-ray centroid not from the X-ray peak, we estimated the position of the X-ray peak also for the \HIFLUGCS\ clusters using the procedure described in Sec. \ref{sec:xray} and we used them to measure $D_{X,BCG}$, that we normalized using their $R_{500}$ estimate.
 The offset distribution shown in the left column of Fig. \ref{fig:xray_comp} has a median value $3.8\ 10^{-3}$ and IQR $1.8\ 10^{-2}$.
\item REXCESS \citep{bori07} is a representative and statistically unbiased subsample of 33 galaxy clusters extracted from the {\it REFLEX} cluster catalogue with a rigorous selection in the luminosity-redshift space (see details in \citealt{bori07}). 
 The BCG coordinates, the offset from the X-ray peak and $R_{500}$ have been published by \citet{haarsma10} for 30 objects. The BCG is estimated basing on optical data obtained with instruments with field of view about $5-7$ arcmin across, which can be smaller than $R_{500}$ of the clusters for a large part of the sample. It is thus possible that some of the offsets may be underestimated. The $D_{X,BCG}$ distribution shown in the middle column of Fig. \ref{fig:xray_comp} has a median value $7.9\ 10^{-3}$ and IQR $1.0\ 10^{-2}$.
\item MACS \citep{ebeling01} is a survey to find the most massive clusters at high redshift $z>0.3$ starting from the RASS catalogue and using optical data to confirm cluster candidates. The offset between the X-ray peak and the BCG has been measured by \citet{mann_ebe} for a subsample of 108 objects, starting from a flux threshold ($F_X[0.1-2.4\ \rm{keV}] > 10^{-12}\ \rm{erg\ s}^{-1}\rm{cm}^{-2}$) with additional luminosity and redshift criteria ($L_X[0.1-2.4\ \rm{keV}] > 5 \ 10^{44}\ \rm{erg\ s}^{-1}$ and $z>0.15$). Since the authors do not provide the $R_{500}$ values for their clusters but provide the X-ray luminosity, we estimated $M_{500}$ and $R_{500}$ using the $L-M$ scaling relation by \citet{pratt09}. The BCG is estimated for 77 clusters basing on imaging data from the UH2.2m telescope, with a field of view of $7.5\times7.5$ arcmin, which is larger than the $R_{500}$ region for the majority of these clusters, but not for all of them. For the remaining clusters, the BCG was estimated basing on SDSS or DSS data but the searching radius is not specified. It is thus possible that some of the offsets may be underestimated.
The $D_{X,BCG}$ distribution shown in the right column of Fig. \ref{fig:xray_comp} has a median value $8.7\ 10^{-3}$ and IQR $1.7\ 10^{-2}$.
\end{itemize}
As shown in Fig. \ref{fig:boxplot}, the three X-ray samples feature different redshift and mass distributions. While \HIFLUGCS\ is composed mainly of local and relatively low-mass objects, MACS by construction contains massive systems at high redshift. REXCESS contains objects at intermediate redshift, with median mass similar to \HIFLUGCS . \\
In Fig.  \ref{fig:xray_comp} we compare the distribution of $D_{X,BCG}$ in units of $R_{500}$ of our \Planck\ sample with the three X-ray samples described above, and show the normalized histogram,  and the cumulative distribution. In all cases, we note that the \Planck\ distribution is skewed towards larger offset than the X-ray ones and that the \Planck\ cumulative distribution rises less steeply than for the X-ray samples. 
The visual impression that the distributions are different, is supported also by the differences in the medians ($0.017$ vs $4-8\ 10^{-3}\ R_{500}$)  and IQR ($0.066$ vs $0.01-0.04\ R_{500}$). \\
We applied the same statistical test as in Sec. \ref{sec:distr} to assess the probability that each of the X-ray selected samples may be drawn from the same parent distribution of our \Planck\ sample and we report the results in Table \ref{table:ks}.
 The significance of the results depends on the tests applied and on the samples: the null hypothesis probability is always $< 0.4\%$ for the MACS and HIFLUGCS sample and of the order of $2-4\%$ for the smaller REXCESS sample. We can thus conclude with a high reliability that the offset distribution in the  \Planck\ sample is different than in the X-ray selected samples. \\
One possible concern in comparing the normalized distributions of the \Planck\ and X-ray samples is a possible difference in the $R_{500}$ estimate. To test this, we compared  the $R_{500}$ values for clusters in common with the \Planck\ sample (28 objects in \HIFLUGCS , 7 in REXCESS and 32 in MACS). The points show a $\sim 10\%$ scatter around the equality line, which likely reflects the scatter in the parent scaling relations used to estimate $R_{500}$, and a small systematic offset, with \Planck\ $R_{500}$ values being on average larger by $3\%$ than $R_{500}$ values in X-ray samples. This means that offsets in the \Planck\ sample are on average slightly smaller than the offsets in X-ray samples for common clusters. Correcting for this small systematic effect would thus lead to larger offsets in \Planck\ clusters, making the SZ offset distribution even more different with respect to the X-ray ones and therefore a systematic bias in the $R_{500}$ estimate cannot be used to explain the discrepancy we found. \\ 
Following our classification scheme ($D_{X,BCG}<0.02 R_{500}$, Sec. \ref{sec:distr}), we calculated the fraction of relaxed objects to be 
$(74\pm 5)\%$ in \HIFLUGCS\, $(73\pm 4) \%$ in MACS and $(77\pm 7)\%$ in REXCESS, while it is only $(52\pm 4)\%$ in our sample.
We computed with a MonteCarlo simulation the probability of obtaining randomly from the \Planck\ sample the fraction of relaxed clusters of X-ray samples and found 
$0.05\%$ for {\it HIFLUGCS}, $<0.001\%$ for MACS, $0.2\%$ for REXCESS. We conclude that 
the fraction of relaxed objects is significantly larger in X-ray samples than in the \Planck\ SZ-selected sample. \\

\section{Discussion}
\label{sec:discussion}
\subsection{SZ vs X-ray selection}
\label{sec:sz_vs_x}
The analysis of the distribution of $D_{X,BCG}$ and the comparison with X-ray selected samples shown in Sec. \ref{sec:results}, allow us to address the question we asked in Sec. \ref{sec:intro}.  Indeed, we can answer that the distribution of our indicator is significantly different in the \Planck\ sample with respect to all X-ray selected samples we considered. The significance of this result can be assessed both with statistical tests on the whole distributions and on the fraction of relaxed clusters. In the former test, the null hypothesis probability that the Planck sample and each of the X-ray selected samples are drawn from the same parent distribution is always $< 0.4\%$ (depending on the test) for MACS and \HIFLUGCS\ and of the order of $2-4\%$ for the smaller REXCESS sample. In the latter comparison, the fraction of relaxed objects in the \Planck\ sample ($52\pm4\%$) differs at more than $3\sigma$ from the fraction in X-ray selected samples ($3.4\sigma$ HIFLUGCS, $3.7\sigma$ MACS and $3.1\sigma$ REXCESS, where $\sigma$ is the combined uncertainty obtained by adding in quadrature the errors in each data set). Therefore, we can answer that, according to our indicator $D_{X,BCG}$, the dynamical state of \Planck\ SZ selected clusters is significantly different from X-ray selected samples.\\
We now address the origin of this result, which may be due either to different selection effects in SZ versus X-ray surveys or to a different mass and redshift distribution in the \Planck\ and in X-ray samples (Sec. \ref{sec:distr}).
The three X-ray selected samples we considered feature different properties, reflecting their selection functions: while \HIFLUGCS\ is mainly composed of local and relatively low-mass systems, clusters in MACS are massive systems at $z>0.1$ and the REXCESS sample shows intermediate properties. \Planck\  clusters are mainly massive objects ($2-20\ 10^{14}M_\odot$) with a broad redshift distribution. The fact that we find similar results when comparing the \Planck\ sample both with a local low-mass sample as \HIFLUGCS\ and with a high-mass high$-z$ sample as MACS suggests that the differences we found are likely not due to the different mass and redshift distributions, but rather to different selection effects. To make a further test, we compared our high$-z$ and high-mass \Planck\ subsample (Sec. \ref{sec:distr}) with subsamples extracted from the MACS sample with the same criteria ($z>0.16$, basically the whole MACS, and $M_{500}>6.4\times 10^{14}M_\odot$) and we applied the KS and MWW$-U$ tests. We still find significant differences, with null-hypothesis probabilities $p_0<1\%$ (Table \ref{table:ks}) suggesting that a large part of the discrepancy is due to the selection method. \\
It is well known that X-ray selection is biased towards relaxed clusters with a centrally peaked surface brightness profile(``cool core'' clusters, or CC): \citet{eckert11} estimate that the fraction of strong CC clusters in HIFLUGCS is overestimated by $29\%$, correct for this bias and predict this fraction to be in the range $35-37\%$. While the fraction of CC objects slightly depends on the indicator used to classify clusters, the fraction reported above is much lower than the value reported for most X-ray selected samples. 
The offset between the X-ray peak and the BCG is not a direct indicator of a cool core, although it has been shown to correlate well with the core state \citep{sanderson09} and we cannot use it to make a direct comparison between our fraction of relaxed objects and the CC fraction reported above. 
However, \citet{eckert11} provided an ``unbiased'' subsample of HIFLUGCS, which should be free of the CC bias, and we could estimate the fraction of relaxed objects in this subsample using the offsets measured by \citet{zhang11}. We found a relaxed fraction of $68 \pm 7$\% in the HIFLUGCS-unbiased subsample, which is smaller than in the full HIFLUGCS sample but
still significantly larger that the value in the \Planck\ sample. This residual discrepancy may results from several factors. First of all, the HIFLUGCS-unbiased subsample is not complete, as it was built as a subset of the HIFLUGCS sample and not of the parent RASS data (see discussion in \citealt{eckert11}). 
More importantly, HIFLUGCS and its unbiased subsample have a very different mass and redshift distribution (Fig. \ref{fig:boxplot}) with respect to the \Planck\ sample, extending to lower masses and redshift. As shown in Sec. \ref{sec:distr}, we tend to find larger relaxed fractions in low-mass and low-redshift samples. Finally, it is also possible that the \Planck\ sample may be biased in the opposite direction of X-ray surveys, by preferentially selecting disturbed objects, but it is not possible to separate those effects with present data.

\subsection{Comparison with previous SZ results}
A first attempt to characterize the dynamical state of SZ-selected clusters has been performed by \citet{song12} for the the first 720 deg$^2$ survey of the South Pole Telescope. They use the offset between the BCG and the SZ centroid as indicator of dynamical state and compare the distribution in their sample with the distribution of the BCG-X ray peak offset for other X-ray based samples, namely \citet{lin_mohr} and \citet{mann_ebe}. They report a good agreement between their observed distribution and the \citet{lin_mohr} sample ($41\%$ probability of consistency), while the agreement is not good with the \citet{mann_ebe} sample ($0.46\%$ null hypothesis probability). They justify this disagreement in terms of differences in the BCG selection procedure and decide to compare their results only with the more consistent \citet{lin_mohr} sample, concluding that there is no compelling evidence that the dynamical state of SZ selected clusters is different than in X-ray selected clusters. \\
However, the number of objects where the difference in the BCG selection procedure between the \citet{mann_ebe} procedure and the SPT one may have lead to a different measurement of the offset is very limited (2-3 clusters, Mann, private communication).
These include cases where: i) two or more elliptical galaxies with colors consistent with the clusters and similar magnitudes lie within the virial radius of the cluster; ii) double clusters.
The exclusion of the MACS sample from the comparison is thus not justified as this small number of objects may not have influenced the properties of the whole distribution.
More importantly, we underline that the \citet{mann_ebe} sample is a well-defined X-ray selected sample, while the \citet{lin_mohr} is not X-ray selected: it is an archival sample built from a collection of X-ray cluster catalogs with published temperature and with a redshift cut $z<0.09$. Finally, the position of the SZ centroid does not necessarily coincide with the position of the X-ray peak and therefore the comparison of the SPT distribution with X-ray samples is not straightforward. \\
Another important result on the properties of clusters selected by SPT through the SZ effect has been published by \citet{mcdonald}, who analyzed the \chandra\ observations of the highest S/N detections in the SPT survey. While the main objective of their paper is the evolution of the core properties with time, \citet{mcdonald} also measure the fraction of cool cores in their total sample and found it to be in the range $10-40\%$. While its exact value depends on the indicator and evolves with redshift
 the fraction of cool core is in any case smaller than the typical values observed in X-ray selected samples.  As discussed in Sec. \ref{sec:sz_vs_x}, the CC fraction cannot be directly related to our relaxed fraction, as measured by our dynamical indicator. Nonetheless, he result in \citet{mcdonald} provides an independent indirect indication of different selection effects between SZ and X-ray surveys.
\section{Summary and conclusions}
In this paper, we studied the dynamical state of a representative subsample of the catalogue of galaxy clusters observed by \Planck\ with the SZ-effect \citep{PSZ1}. We have used as indicator of dynamical state the projected offset between the position of the X-ray peak and the position of the BCG, which is expected to be small for relaxed objects and larger for disturbed systems. By dividing our sample in redshift and mass bins, we find a suggestive indication (at $2.5-2.8\sigma$) that high-mass and high-redshift subsamples host more disturbed objects than the low-mass and low-z samples. 
We compared the distributions of our indicator in the \Planck\ sample with three X-ray selected catalogues (\HIFLUGCS , MACS and REXCESS) and found that the distributions are significantly different: the fraction of relaxed objects in our sample is significantly smaller ($>3\sigma$) with respect to the X-ray samples and the statistical test we applied to the $D_{X-BCG}$ distributions return very small probabilities that the \Planck\ and X-ray samples are drawn from the same parent distribution. We have shown that this difference is not due to the mass and redshift distributions, but is likely due to different selection effects affecting X-rays (the so-called ``cool core bias'') and, possibly, SZ surveys. Indeed, we confirm with our analysis the early impression that many \Planck\ detected clusters are dynamically disturbed systems \citep{planck_early_IX} and we provide the first observational indication that the SZ-selection is less biased towards relaxed objects than the X-ray selection. \\
An intrinsic limitation of the indicator used in our analysis is that it suffers from projection effects: if a merger is separating the BCG and the X-ray peak mainly along the line of sight, $D_{X-BCG}$ would be underestimated with respect to the true physical offset. Consequently a number of dynamically disturbed objects are mis-classified as relaxed and the fraction of relaxed objects measured with $D_{X-BCG}$ in all samples is likely overestimated. This effect should be taken into account when comparing it to similar quantities obtained with other indicators, which do not suffer from projection effects. We tried to correct for this effect (as described in Appendix \ref{app_B}) and we estimate the real fraction of relaxed objects in the \Planck\ sample to be $45\%$. However, even after this correction, $D_{X-BCG}$ is still a dynamical indicator and the comparison of our relaxed fraction with the cool core fraction obtained from thermodynamical indicators is not straightforward and requires several assumptions. 
 A further limitation of our analysis is our estimate of the BCG from a heterogeneous set of literature information which may result in an underestimation of the measured offset for a few clusters (Sec.\ref{sec:BCG}). This limitation does not affect only our sample but to a different extent also the MACS and REXCESS sample (Sec. \ref{sec:xsamples}). \\
 Our work should be considered as a first step towards a description of the dynamical and thermodynamical state of \Planck\ SZ-selected clusters. It will soon be possible to complement it and verify these results with several morphological indicators on X-ray images (center shift, power ratios, concentration parameter) as well as thermodynamical quantities (central entropy and cooling time, entropy ratio) on similar representative subsamples of the \Planck\ catalogue. These studies will allow us to firmly assess the fraction of cool-cores in \Planck\ SZ-selected samples and compare them with the values derived from X-ray surveys and with predictions of simulations and thus establish the difference between SZ and X-ray selected surveys.

\section*{Acknowledgements}
We thank H. Ebeling for providing the BCG coordinates of 4 MACS clusters. MR acknowledges useful discussions with A. Mann, J.B. Melin \& S. Tunesi. FG acknowledges the financial contribution from contracts ASI-INAF I/037/12/0 and PRIN-INAF 2012 ``A unique dataset to address the most compelling open questions about X-ray clusters''.   This research has made use of the NASA/IPAC Extragalactic Database (NED) which is operated by the Jet Propulsion Laboratory, California Institute of Technology, under contract with the National Aeronautics and Space Administration.

\bibliographystyle{mnras}
\bibliography{mybib}

\begin{thebibliography}{}
\makeatletter
\relax
\def\mn@urlcharsother{\let\do\@makeother \do\$\do\&\do\#\do\^\do\_\do\%\do\~}
\def\mn@doi{\begingroup\mn@urlcharsother \@ifnextchar [ {\mn@doi@}
  {\mn@doi@[]}}
\def\mn@doi@[#1]#2{\def\@tempa{#1}\ifx\@tempa\@empty \href
  {http://dx.doi.org/#2} {doi:#2}\else \href {http://dx.doi.org/#2} {#1}\fi
  \endgroup}
\def\mn@eprint#1#2{\mn@eprint@#1:#2::\@nil}
\def\mn@eprint@arXiv#1{\href {http://arxiv.org/abs/#1} {{\tt arXiv:#1}}}
\def\mn@eprint@dblp#1{\href {http://dblp.uni-trier.de/rec/bibtex/#1.xml}
  {dblp:#1}}
\def\mn@eprint@#1:#2:#3:#4\@nil{\def\@tempa {#1}\def\@tempb {#2}\def\@tempc
  {#3}\ifx \@tempc \@empty \let \@tempc \@tempb \let \@tempb \@tempa \fi \ifx
  \@tempb \@empty \def\@tempb {arXiv}\fi \@ifundefined
  {mn@eprint@\@tempb}{\@tempb:\@tempc}{\expandafter \expandafter \csname
  mn@eprint@\@tempb\endcsname \expandafter{\@tempc}}}

\bibitem[\protect\citeauthoryear{{Battaglia}, {Bond}, {Pfrommer}  \&
  {Sievers}}{{Battaglia} et~al.}{2012}]{battaglia12}
{Battaglia} N.,  {Bond} J.~R.,  {Pfrommer} C.,   {Sievers} J.~L.,  2012,
  \mn@doi [\apj] {10.1088/0004-637X/758/2/74}, \href
  {http://cdsads.u-strasbg.fr/abs/2012ApJ...758...74B} {758, 74}

\bibitem[\protect\citeauthoryear{{Beers} \& {Geller}}{{Beers} \&
  {Geller}}{1983}]{Beers.ea:83}
{Beers} T.~C.,  {Geller} M.~J.,  1983, \mn@doi [\apj] {10.1086/161463}, \href
  {http://adsabs.harvard.edu/abs/1983ApJ...274..491B} {274, 491}

\bibitem[\protect\citeauthoryear{{Benson} et~al.,}{{Benson}
  et~al.}{2013}]{benson13}
{Benson} B.~A.,  et~al., 2013, \mn@doi [\apj] {10.1088/0004-637X/763/2/147},
  \href {http://cdsads.u-strasbg.fr/abs/2013ApJ...763..147B} {763, 147}

\bibitem[\protect\citeauthoryear{{Bildfell}, {Hoekstra}, {Babul}  \&
  {Mahdavi}}{{Bildfell} et~al.}{2008}]{bildfell08}
{Bildfell} C.,  {Hoekstra} H.,  {Babul} A.,   {Mahdavi} A.,  2008, \mn@doi
  [\mnras] {10.1111/j.1365-2966.2008.13699.x}, \href
  {http://cdsads.u-strasbg.fr/abs/2008MNRAS.389.1637B} {389, 1637}

\bibitem[\protect\citeauthoryear{{Bleem} et~al.,}{{Bleem}
  et~al.}{2015}]{SPT_cat}
{Bleem} L.~E.,  et~al., 2015, \mn@doi [\apjs] {10.1088/0067-0049/216/2/27},
  \href {http://cdsads.u-strasbg.fr/abs/2015ApJS..216...27B} {216, 27}

\bibitem[\protect\citeauthoryear{{B{\"o}hringer} et~al.,}{{B{\"o}hringer}
  et~al.}{2007}]{bori07}
{B{\"o}hringer} H.,  et~al., 2007, \mn@doi [\aap] {10.1051/0004-6361:20066740},
  \href {http://cdsads.u-strasbg.fr/abs/2007A%26A...469..363B} {469, 363}

\bibitem[\protect\citeauthoryear{{Bonafede}, {Intema}, {Br{\"u}ggen},
  {Girardi}, {Nonino}, {Kantharia}, {van Weeren}  \&
  {R{\"o}ttgering}}{{Bonafede} et~al.}{2014}]{bonafede14}
{Bonafede} A.,  {Intema} H.~T.,  {Br{\"u}ggen} M.,  {Girardi} M.,  {Nonino} M.,
   {Kantharia} N.,  {van Weeren} R.~J.,   {R{\"o}ttgering} H.~J.~A.,  2014,
  \mn@doi [\apj] {10.1088/0004-637X/785/1/1}, \href
  {http://cdsads.u-strasbg.fr/abs/2014ApJ...785....1B} {785, 1}

\bibitem[\protect\citeauthoryear{{Boschin}, {Girardi}, {Barrena}  \&
  {Nonino}}{{Boschin} et~al.}{2012}]{boschin12}
{Boschin} W.,  {Girardi} M.,  {Barrena} R.,   {Nonino} M.,  2012, \mn@doi
  [\aap] {10.1051/0004-6361/201118076}, \href
  {http://cdsads.u-strasbg.fr/abs/2012A%26A...540A..43B} {540, A43}

\bibitem[\protect\citeauthoryear{{Brough}, {Collins}, {Burke}, {Lynam}  \&
  {Mann}}{{Brough} et~al.}{2005}]{Brough.ea:05}
{Brough} S.,  {Collins} C.~A.,  {Burke} D.~J.,  {Lynam} P.~D.,   {Mann} R.~G.,
  2005, \mn@doi [\mnras] {10.1111/j.1365-2966.2005.09679.x}, \href
  {http://adsabs.harvard.edu/abs/2005MNRAS.364.1354B} {364, 1354}

\bibitem[\protect\citeauthoryear{{Brough}, {Couch}, {Collins}, {Jarrett},
  {Burke}  \& {Mann}}{{Brough} et~al.}{2008}]{Brough.ea:08}
{Brough} S.,  {Couch} W.~J.,  {Collins} C.~A.,  {Jarrett} T.,  {Burke} D.~J.,
  {Mann} R.~G.,  2008, \mn@doi [\mnras] {10.1111/j.1745-3933.2008.00442.x},
  \href {http://adsabs.harvard.edu/abs/2008MNRAS.385L.103B} {385, L103}

\bibitem[\protect\citeauthoryear{{Buote} \& {Tsai}}{{Buote} \&
  {Tsai}}{1995}]{buotetsai}
{Buote} D.~A.,  {Tsai} J.~C.,  1995, \mn@doi [\apj] {10.1086/176326}, \href
  {http://adsabs.harvard.edu/abs/1995ApJ...452..522B} {452, 522}

\bibitem[\protect\citeauthoryear{{Cavagnolo}, {Donahue}, {Voit}  \&
  {Sun}}{{Cavagnolo} et~al.}{2009}]{cava09}
{Cavagnolo} K.~W.,  {Donahue} M.,  {Voit} G.~M.,   {Sun} M.,  2009, \mn@doi
  [\apjs] {10.1088/0067-0049/182/1/12}, \href
  {http://adsabs.harvard.edu/abs/2009ApJS..182...12C} {182, 12}

\bibitem[\protect\citeauthoryear{{Coziol}, {Andernach}, {Caretta},
  {Alamo-Mart{\'{\i}}nez}  \& {Tago}}{{Coziol} et~al.}{2009}]{coziol09}
{Coziol} R.,  {Andernach} H.,  {Caretta} C.~A.,  {Alamo-Mart{\'{\i}}nez} K.~A.,
    {Tago} E.,  2009, \mn@doi [\aj] {10.1088/0004-6256/137/6/4795}, \href
  {http://cdsads.u-strasbg.fr/abs/2009AJ....137.4795C} {137, 4795}

\bibitem[\protect\citeauthoryear{{Crawford}, {Edge}, {Fabian}, {Allen},
  {Bohringer}, {Ebeling}, {McMahon}  \& {Voges}}{{Crawford}
  et~al.}{1995}]{crawford95}
{Crawford} C.~S.,  {Edge} A.~C.,  {Fabian} A.~C.,  {Allen} S.~W.,  {Bohringer}
  H.,  {Ebeling} H.,  {McMahon} R.~G.,   {Voges} W.,  1995, \mnras, \href
  {http://cdsads.u-strasbg.fr/abs/1995MNRAS.274...75C} {274, 75}

\bibitem[\protect\citeauthoryear{{Crawford}, {Allen}, {Ebeling}, {Edge}  \&
  {Fabian}}{{Crawford} et~al.}{1999}]{crawford99}
{Crawford} C.~S.,  {Allen} S.~W.,  {Ebeling} H.,  {Edge} A.~C.,   {Fabian}
  A.~C.,  1999, \mn@doi [\mnras] {10.1046/j.1365-8711.1999.02583.x}, \href
  {http://cdsads.u-strasbg.fr/abs/1999MNRAS.306..857C} {306, 857}

\bibitem[\protect\citeauthoryear{{Ebeling}, {Edge}  \& {Henry}}{{Ebeling}
  et~al.}{2001}]{ebeling01}
{Ebeling} H.,  {Edge} A.~C.,   {Henry} J.~P.,  2001, \mn@doi [\apj]
  {10.1086/320958}, \href {http://cdsads.u-strasbg.fr/abs/2001ApJ...553..668E}
  {553, 668}

\bibitem[\protect\citeauthoryear{{Ebeling}, {Ma}, {Kneib}, {Jullo}, {Courtney},
  {Barrett}, {Edge}  \& {Le Borgne}}{{Ebeling} et~al.}{2009}]{ebeling09}
{Ebeling} H.,  {Ma} C.~J.,  {Kneib} J.-P.,  {Jullo} E.,  {Courtney} N.~J.~D.,
  {Barrett} E.,  {Edge} A.~C.,   {Le Borgne} J.-F.,  2009, \mn@doi [\mnras]
  {10.1111/j.1365-2966.2009.14502.x}, \href
  {http://cdsads.u-strasbg.fr/abs/2009MNRAS.395.1213E} {395, 1213}

\bibitem[\protect\citeauthoryear{{Eckert}, {Molendi}  \& {Paltani}}{{Eckert}
  et~al.}{2011}]{eckert11}
{Eckert} D.,  {Molendi} S.,   {Paltani} S.,  2011, \mn@doi [\aap]
  {10.1051/0004-6361/201015856}, \href
  {http://cdsads.u-strasbg.fr/abs/2011A%26A...526A..79E} {526, A79}

\bibitem[\protect\citeauthoryear{{Edge}}{{Edge}}{1991}]{Edge:91}
{Edge} A.~C.,  1991, \mnras, \href
  {http://adsabs.harvard.edu/abs/1991MNRAS.250..103E} {250, 103}

\bibitem[\protect\citeauthoryear{{Edge} \& {Stewart}}{{Edge} \&
  {Stewart}}{1991}]{Edge.ea:91}
{Edge} A.~C.,  {Stewart} G.~C.,  1991, \mnras, \href
  {http://adsabs.harvard.edu/abs/1991MNRAS.252..414E} {252, 414}

\bibitem[\protect\citeauthoryear{{Evans} et~al.,}{{Evans}
  et~al.}{2010}]{evans10}
{Evans} I.~N.,  et~al., 2010, \mn@doi [\apjs] {10.1088/0067-0049/189/1/37},
  \href {http://cdsads.u-strasbg.fr/abs/2010ApJS..189...37E} {189, 37}

\bibitem[\protect\citeauthoryear{Feigelson \& Babu}{Feigelson \&
  Babu}{2012}]{feigelson2012}
Feigelson E.,  Babu G.,  2012, Modern Statistical Methods for Astronomy: With R
  Applications.
Cambridge University Press, \url
  {https://books.google.it/books?id=M6O1yxpvf2gC}

\bibitem[\protect\citeauthoryear{Gradshteyn \& Ryzhik}{Gradshteyn \&
  Ryzhik}{2007}]{gradshteyn2007}
Gradshteyn I.~S.,  Ryzhik I.~M.,  2007, Table of integrals, series, and
  products, seventh edn.
Elsevier/Academic Press, Amsterdam

\bibitem[\protect\citeauthoryear{{Guzzo} et~al.,}{{Guzzo}
  et~al.}{2009}]{guzzo09}
{Guzzo} L.,  et~al., 2009, \mn@doi [\aap] {10.1051/0004-6361/200810838}, \href
  {http://cdsads.u-strasbg.fr/abs/2009A%26A...499..357G} {499, 357}

\bibitem[\protect\citeauthoryear{{Haarsma} et~al.,}{{Haarsma}
  et~al.}{2010}]{haarsma10}
{Haarsma} D.~B.,  et~al., 2010, \mn@doi [\apj] {10.1088/0004-637X/713/2/1037},
  \href {http://cdsads.u-strasbg.fr/abs/2010ApJ...713.1037H} {713, 1037}

\bibitem[\protect\citeauthoryear{{Hashimoto}, {Henry}  \&
  {Boehringer}}{{Hashimoto} et~al.}{2014}]{Hashimoto.ea:14}
{Hashimoto} Y.,  {Henry} J.~P.,   {Boehringer} H.,  2014, \mn@doi [\mnras]
  {10.1093/mnras/stu311}, \href
  {http://adsabs.harvard.edu/abs/2014MNRAS.440..588H} {440, 588}

\bibitem[\protect\citeauthoryear{{Hasselfield} et~al.,}{{Hasselfield}
  et~al.}{2013}]{ACT_cat}
{Hasselfield} M.,  et~al., 2013, \mn@doi [\jcap]
  {10.1088/1475-7516/2013/07/008}, \href
  {http://adsabs.harvard.edu/abs/2013JCAP...07..008H} {7, 8}

\bibitem[\protect\citeauthoryear{{Hoffer}, {Donahue}, {Hicks}  \&
  {Barthelemy}}{{Hoffer} et~al.}{2012}]{hoffer12}
{Hoffer} A.~S.,  {Donahue} M.,  {Hicks} A.,   {Barthelemy} R.~S.,  2012,
  \mn@doi [\apjs] {10.1088/0067-0049/199/1/23}, \href
  {http://cdsads.u-strasbg.fr/abs/2012ApJS..199...23H} {199, 23}

\bibitem[\protect\citeauthoryear{{Hudson}, {Mittal}, {Reiprich}, {Nulsen},
  {Andernach}  \& {Sarazin}}{{Hudson} et~al.}{2010}]{hudson10}
{Hudson} D.~S.,  {Mittal} R.,  {Reiprich} T.~H.,  {Nulsen} P.~E.~J.,
  {Andernach} H.,   {Sarazin} C.~L.,  2010, \mn@doi [\aap]
  {10.1051/0004-6361/200912377}, \href
  {http://adsabs.harvard.edu/abs/2010A%26A...513A..37H} {513, A37}

\bibitem[\protect\citeauthoryear{{Jones} \& {Forman}}{{Jones} \&
  {Forman}}{1984}]{JonesC.ea:84}
{Jones} C.,  {Forman} W.,  1984, \mn@doi [\apj] {10.1086/161591}, \href
  {http://adsabs.harvard.edu/abs/1984ApJ...276...38J} {276, 38}

\bibitem[\protect\citeauthoryear{{Jones} \& {Forman}}{{Jones} \&
  {Forman}}{1999}]{JonesC.ea:99}
{Jones} C.,  {Forman} W.,  1999, \mn@doi [\apj] {10.1086/306646}, \href
  {http://adsabs.harvard.edu/cgi-bin/nph-bib_query?bibcode=1999ApJ...511...65J&db_key=AST}
  {511, 65}

\bibitem[\protect\citeauthoryear{{Katayama}, {Hayashida}, {Takahara}  \&
  {Fujita}}{{Katayama} et~al.}{2003}]{Katayama.ea:03}
{Katayama} H.,  {Hayashida} K.,  {Takahara} F.,   {Fujita} Y.,  2003, \mn@doi
  [\apj] {10.1086/346126}, \href
  {http://adsabs.harvard.edu/abs/2003ApJ...585..687K} {585, 687}

\bibitem[\protect\citeauthoryear{{Koester} et~al.,}{{Koester}
  et~al.}{2007}]{maxbcg}
{Koester} B.~P.,  et~al., 2007, \mn@doi [\apj] {10.1086/509599}, \href
  {http://cdsads.u-strasbg.fr/abs/2007ApJ...660..239K} {660, 239}

\bibitem[\protect\citeauthoryear{{Krause}, {Pierpaoli}, {Dolag}  \&
  {Borgani}}{{Krause} et~al.}{2012}]{krause12}
{Krause} E.,  {Pierpaoli} E.,  {Dolag} K.,   {Borgani} S.,  2012, \mn@doi
  [\mnras] {10.1111/j.1365-2966.2011.19844.x}, \href
  {http://cdsads.u-strasbg.fr/abs/2012MNRAS.419.1766K} {419, 1766}

\bibitem[\protect\citeauthoryear{{Leccardi}, {Rossetti}  \&
  {Molendi}}{{Leccardi} et~al.}{2010}]{leccardi10}
{Leccardi} A.,  {Rossetti} M.,   {Molendi} S.,  2010, \mn@doi [\aap]
  {10.1051/0004-6361/200913094}, \href
  {http://adsabs.harvard.edu/abs/2010A%26A...510A..82L} {510, A82}

\bibitem[\protect\citeauthoryear{{Lin} \& {Mohr}}{{Lin} \&
  {Mohr}}{2004}]{lin_mohr}
{Lin} Y.-T.,  {Mohr} J.~J.,  2004, \mn@doi [\apj] {10.1086/425412}, \href
  {http://cdsads.u-strasbg.fr/abs/2004ApJ...617..879L} {617, 879}

\bibitem[\protect\citeauthoryear{{Lin}, {McDonald}, {Benson}  \&
  {Miller}}{{Lin} et~al.}{2015}]{lin15}
{Lin} H.~W.,  {McDonald} M.,  {Benson} B.,   {Miller} E.,  2015, \mn@doi [\apj]
  {10.1088/0004-637X/802/1/34}, \href
  {http://cdsads.u-strasbg.fr/abs/2015ApJ...802...34L} {802, 34}

\bibitem[\protect\citeauthoryear{{Malumuth}, {Kriss}, {Dixon}, {Ferguson}  \&
  {Ritchie}}{{Malumuth} et~al.}{1992}]{Malumuth.ea:92}
{Malumuth} E.~M.,  {Kriss} G.~A.,  {Dixon} W.~V.~D.,  {Ferguson} H.~C.,
  {Ritchie} C.,  1992, \mn@doi [\aj] {10.1086/116250}, \href
  {http://adsabs.harvard.edu/abs/1992AJ....104..495M} {104, 495}

\bibitem[\protect\citeauthoryear{{Mann} \& {Ebeling}}{{Mann} \&
  {Ebeling}}{2012}]{mann_ebe}
{Mann} A.~W.,  {Ebeling} H.,  2012, \mn@doi [\mnras]
  {10.1111/j.1365-2966.2011.20170.x}, \href
  {http://adsabs.harvard.edu/abs/2012MNRAS.420.2120M} {420, 2120}

\bibitem[\protect\citeauthoryear{{McDonald}, {Benson}, {Vikhlinin}, {Stalder},
  {Bleem}, {de Haan}, {Lin}  \& {Aird}}{{McDonald} et~al.}{2013}]{mcdonald}
{McDonald} M.,  {Benson} B.~A.,  {Vikhlinin} A.,  {Stalder} B.,  {Bleem} L.~E.,
   {de Haan} T.,  {Lin} H.~W.,   {Aird} K.~A.,  2013, \mn@doi [\apj]
  {10.1088/0004-637X/774/1/23}, \href
  {http://cdsads.u-strasbg.fr/abs/2013ApJ...774...23M} {774, 23}

\bibitem[\protect\citeauthoryear{{McNamara} et~al.,}{{McNamara}
  et~al.}{2006}]{mcnamara06}
{McNamara} B.~R.,  et~al., 2006, \mn@doi [\apj] {10.1086/505859}, \href
  {http://adsabs.harvard.edu/abs/2006ApJ...648..164M} {648, 164}

\bibitem[\protect\citeauthoryear{{Menanteau} et~al.,}{{Menanteau}
  et~al.}{2010}]{menanteau10}
{Menanteau} F.,  et~al., 2010, \mn@doi [\apj] {10.1088/0004-637X/723/2/1523},
  \href {http://cdsads.u-strasbg.fr/abs/2010ApJ...723.1523M} {723, 1523}

\bibitem[\protect\citeauthoryear{{Motl}, {Hallman}, {Burns}  \&
  {Norman}}{{Motl} et~al.}{2005}]{motl05}
{Motl} P.~M.,  {Hallman} E.~J.,  {Burns} J.~O.,   {Norman} M.~L.,  2005,
  \mn@doi [\apjl] {10.1086/430144}, \href
  {http://cdsads.u-strasbg.fr/abs/2005ApJ...623L..63M} {623, L63}

\bibitem[\protect\citeauthoryear{{Oegerle} \& {Hill}}{{Oegerle} \&
  {Hill}}{2001}]{Oegerle.ea:01}
{Oegerle} W.~R.,  {Hill} J.~M.,  2001, \mn@doi [\aj] {10.1086/323536}, \href
  {http://adsabs.harvard.edu/abs/2001AJ....122.2858O} {122, 2858}

\bibitem[\protect\citeauthoryear{{Owers}, {Nulsen}, {Couch}  \&
  {Markevitch}}{{Owers} et~al.}{2009}]{owers09}
{Owers} M.~S.,  {Nulsen} P.~E.~J.,  {Couch} W.~J.,   {Markevitch} M.,  2009,
  \mn@doi [\apj] {10.1088/0004-637X/704/2/1349}, \href
  {http://adsabs.harvard.edu/abs/2009ApJ...704.1349O} {704, 1349}

\bibitem[\protect\citeauthoryear{{Owers}, {Randall}, {Nulsen}, {Couch}, {David}
   \& {Kempner}}{{Owers} et~al.}{2011}]{owers11}
{Owers} M.~S.,  {Randall} S.~W.,  {Nulsen} P.~E.~J.,  {Couch} W.~J.,  {David}
  L.~P.,   {Kempner} J.~C.,  2011, \mn@doi [\apj] {10.1088/0004-637X/728/1/27},
  \href {http://cdsads.u-strasbg.fr/abs/2011ApJ...728...27O} {728, 27}

\bibitem[\protect\citeauthoryear{{Patel}, {Maddox}, {Pearce},
  {Arag{\'o}n-Salamanca}  \& {Conway}}{{Patel} et~al.}{2006}]{Patel.ea:06}
{Patel} P.,  {Maddox} S.,  {Pearce} F.~R.,  {Arag{\'o}n-Salamanca} A.,
  {Conway} E.,  2006, \mn@doi [\mnras] {10.1111/j.1365-2966.2006.10510.x},
  \href {http://cdsads.u-strasbg.fr/abs/2006MNRAS.370..851P} {370, 851}

\bibitem[\protect\citeauthoryear{{Peres}, {Fabian}, {Edge}, {Allen},
  {Johnstone}  \& {White}}{{Peres} et~al.}{1998}]{peres98}
{Peres} C.~B.,  {Fabian} A.~C.,  {Edge} A.~C.,  {Allen} S.~W.,  {Johnstone}
  R.~M.,   {White} D.~A.,  1998, \mn@doi [\mnras]
  {10.1046/j.1365-8711.1998.01624.x}, \href
  {http://adsabs.harvard.edu/abs/1998MNRAS.298..416P} {298, 416}

\bibitem[\protect\citeauthoryear{{Pipino} \& {Pierpaoli}}{{Pipino} \&
  {Pierpaoli}}{2010}]{pipino10}
{Pipino} A.,  {Pierpaoli} E.,  2010, \mn@doi [\mnras]
  {10.1111/j.1365-2966.2010.16393.x}, \href
  {http://cdsads.u-strasbg.fr/abs/2010MNRAS.404.1603P} {404, 1603}

\bibitem[\protect\citeauthoryear{{Planck Collaboration}}{{Planck
  Collaboration}}{2015}]{cosmoPSZ2}
{Planck Collaboration} 2015, submitted to \aap, arXiv:1502.01597

\bibitem[\protect\citeauthoryear{{Planck Collaboration XX}}{{Planck
  Collaboration XX}}{2014}]{cosmoPSZ1}
{Planck Collaboration XX} 2014, \mn@doi [\aap] {10.1051/0004-6361/201321521},
  \href {http://cdsads.u-strasbg.fr/abs/2014A%26A...571A..20P} {571, A20}

\bibitem[\protect\citeauthoryear{{Planck Collaboration XXIX}}{{Planck
  Collaboration XXIX}}{2014}]{PSZ1}
{Planck Collaboration XXIX} 2014, \mn@doi [\aap] {10.1051/0004-6361/201321523},
  \href {http://cdsads.u-strasbg.fr/abs/2014A%26A...571A..29P} {571, A29}

\bibitem[\protect\citeauthoryear{{Planck Collaboration XXVII}}{{Planck
  Collaboration XXVII}}{2015}]{PSZ2}
{Planck Collaboration XXVII} 2015, \aap in press

\bibitem[\protect\citeauthoryear{{Planck Collaboration XXXII}}{{Planck
  Collaboration XXXII}}{2015}]{PSZ1_update}
{Planck Collaboration XXXII} 2015, \mn@doi [\aap]
  {10.1051/0004-6361/201525787}, \href
  {http://cdsads.u-strasbg.fr/abs/2015A%26A...581A..14P} {581, A14}

\bibitem[\protect\citeauthoryear{{Planck Collaboration} et~al.,}{{Planck
  Collaboration} et~al.}{2011}]{planck_early_IX}
{Planck Collaboration} et~al., 2011, \mn@doi [\aap]
  {10.1051/0004-6361/201116460}, \href
  {http://adsabs.harvard.edu/abs/2011A%26A...536A...9P} {536, A9}

\bibitem[\protect\citeauthoryear{{Poole}, {Fardal}, {Babul}, {McCarthy},
  {Quinn}  \& {Wadsley}}{{Poole} et~al.}{2006}]{poole06}
{Poole} G.~B.,  {Fardal} M.~A.,  {Babul} A.,  {McCarthy} I.~G.,  {Quinn} T.,
  {Wadsley} J.,  2006, \mn@doi [\mnras] {10.1111/j.1365-2966.2006.10916.x},
  \href {http://adsabs.harvard.edu/abs/2006MNRAS.373..881P} {373, 881}

\bibitem[\protect\citeauthoryear{{Postman} \& {Lauer}}{{Postman} \&
  {Lauer}}{1995}]{postman95}
{Postman} M.,  {Lauer} T.~R.,  1995, \mn@doi [\apj] {10.1086/175245}, \href
  {http://cdsads.u-strasbg.fr/abs/1995ApJ...440...28P} {440, 28}

\bibitem[\protect\citeauthoryear{{Pratt}, {Croston}, {Arnaud}  \&
  {B{\"o}hringer}}{{Pratt} et~al.}{2009}]{pratt09}
{Pratt} G.~W.,  {Croston} J.~H.,  {Arnaud} M.,   {B{\"o}hringer} H.,  2009,
  \mn@doi [\aap] {10.1051/0004-6361/200810994}, \href
  {http://cdsads.u-strasbg.fr/abs/2009A%26A...498..361P} {498, 361}

\bibitem[\protect\citeauthoryear{{R Core Team}}{{R Core Team}}{2015}]{r_cite}
{R Core Team} 2015, R: A Language and Environment for Statistical Computing.
R Foundation for Statistical Computing, Vienna, Austria, \url
  {https://www.R-project.org}

\bibitem[\protect\citeauthoryear{{Rawle} et~al.,}{{Rawle}
  et~al.}{2012}]{rawle12}
{Rawle} T.~D.,  et~al., 2012, \mn@doi [\apj] {10.1088/0004-637X/747/1/29},
  \href {http://cdsads.u-strasbg.fr/abs/2012ApJ...747...29R} {747, 29}

\bibitem[\protect\citeauthoryear{{Reiprich} \& {B{\"o}hringer}}{{Reiprich} \&
  {B{\"o}hringer}}{2002}]{reiprich02}
{Reiprich} T.~H.,  {B{\"o}hringer} H.,  2002, \mn@doi [\apj] {10.1086/338753},
  \href {http://cdsads.u-strasbg.fr/abs/2002ApJ...567..716R} {567, 716}

\bibitem[\protect\citeauthoryear{{Rykoff} et~al.,}{{Rykoff}
  et~al.}{2014}]{rykoff14}
{Rykoff} E.~S.,  et~al., 2014, \mn@doi [\apj] {10.1088/0004-637X/785/2/104},
  \href {http://cdsads.u-strasbg.fr/abs/2014ApJ...785..104R} {785, 104}

\bibitem[\protect\citeauthoryear{{Sanderson}, {Edge}  \& {Smith}}{{Sanderson}
  et~al.}{2009}]{sanderson09}
{Sanderson} A.~J.~R.,  {Edge} A.~C.,   {Smith} G.~P.,  2009, \mn@doi [\mnras]
  {10.1111/j.1365-2966.2009.15214.x}, \href
  {http://adsabs.harvard.edu/abs/2009MNRAS.398.1698S} {398, 1698}

\bibitem[\protect\citeauthoryear{{Santos}, {Rosati}, {Tozzi}, {B{\"o}hringer},
  {Ettori}  \& {Bignamini}}{{Santos} et~al.}{2008}]{santos08}
{Santos} J.~S.,  {Rosati} P.,  {Tozzi} P.,  {B{\"o}hringer} H.,  {Ettori} S.,
  {Bignamini} A.,  2008, \mn@doi [\aap] {10.1051/0004-6361:20078815}, \href
  {http://adsabs.harvard.edu/abs/2008A%26A...483...35S} {483, 35}

\bibitem[\protect\citeauthoryear{{Skibba} \& {Macci{\`o}}}{{Skibba} \&
  {Macci{\`o}}}{2011}]{Skibba.ea:11*1}
{Skibba} R.~A.,  {Macci{\`o}} A.~V.,  2011, \mn@doi [\mnras]
  {10.1111/j.1365-2966.2011.19218.x}, \href
  {http://adsabs.harvard.edu/abs/2011MNRAS.416.2388S} {416, 2388}

\bibitem[\protect\citeauthoryear{{Song} et~al.,}{{Song} et~al.}{2012}]{song12}
{Song} J.,  et~al., 2012, \mn@doi [\apj] {10.1088/0004-637X/761/1/22}, \href
  {http://cdsads.u-strasbg.fr/abs/2012ApJ...761...22S} {761, 22}

\bibitem[\protect\citeauthoryear{{Stanford}, {Eisenhardt}, {Dickinson},
  {Holden}  \& {De Propris}}{{Stanford} et~al.}{2002}]{stanford02}
{Stanford} S.~A.,  {Eisenhardt} P.~R.,  {Dickinson} M.,  {Holden} B.~P.,   {De
  Propris} R.,  2002, \mn@doi [\apjs] {10.1086/340972}, \href
  {http://cdsads.u-strasbg.fr/abs/2002ApJS..142..153S} {142, 153}

\bibitem[\protect\citeauthoryear{{Story} et~al.,}{{Story}
  et~al.}{2011}]{story11}
{Story} K.,  et~al., 2011, \mn@doi [\apjl] {10.1088/2041-8205/735/2/L36}, \href
  {http://cdsads.u-strasbg.fr/abs/2011ApJ...735L..36S} {735, L36}

\bibitem[\protect\citeauthoryear{{Stott}, {Edge}, {Smith}, {Swinbank}  \&
  {Ebeling}}{{Stott} et~al.}{2008}]{stott08}
{Stott} J.~P.,  {Edge} A.~C.,  {Smith} G.~P.,  {Swinbank} A.~M.,   {Ebeling}
  H.,  2008, \mn@doi [\mnras] {10.1111/j.1365-2966.2007.12807.x}, \href
  {http://cdsads.u-strasbg.fr/abs/2008MNRAS.384.1502S} {384, 1502}

\bibitem[\protect\citeauthoryear{{Stott} et~al.,}{{Stott}
  et~al.}{2012}]{stott12}
{Stott} J.~P.,  et~al., 2012, \mn@doi [\mnras]
  {10.1111/j.1365-2966.2012.20764.x}, \href
  {http://cdsads.u-strasbg.fr/abs/2012MNRAS.422.2213S} {422, 2213}

\bibitem[\protect\citeauthoryear{{Sun}}{{Sun}}{2009}]{sun09}
{Sun} M.,  2009, \mn@doi [\apj] {10.1088/0004-637X/704/2/1586}, \href
  {http://cdsads.u-strasbg.fr/abs/2009ApJ...704.1586S} {704, 1586}

\bibitem[\protect\citeauthoryear{{Sunyaev} \& {Zeldovich}}{{Sunyaev} \&
  {Zeldovich}}{1970}]{sun70}
{Sunyaev} R.~A.,  {Zeldovich} Y.~B.,  1970, Comments on Astrophysics and Space
  Physics, \href {http://adsabs.harvard.edu/abs/1970CoASP...2...66S} {2, 66}

\bibitem[\protect\citeauthoryear{Sunyaev \& Zeldovich}{Sunyaev \&
  Zeldovich}{1972}]{sun72}
Sunyaev R.~A.,  Zeldovich Y.~B.,  1972, Comments on Astrophysics and Space
  Physics, 4, 173

\bibitem[\protect\citeauthoryear{{Taylor}}{{Taylor}}{2005}]{topcat}
{Taylor} M.~B.,  2005, in {Shopbell} P.,  {Britton} M.,   {Ebert} R.,  eds,
  Astronomical Society of the Pacific Conference Series Vol. 347, Astronomical
  Data Analysis Software and Systems XIV. p.~29

\bibitem[\protect\citeauthoryear{{Valtchanov}, {Murphy}, {Pierre}, {Hunstead}
  \& {L{\'e}monon}}{{Valtchanov} et~al.}{2002}]{valtchanov02}
{Valtchanov} I.,  {Murphy} T.,  {Pierre} M.,  {Hunstead} R.,   {L{\'e}monon}
  L.,  2002, \mn@doi [\aap] {10.1051/0004-6361:20020940}, \href
  {http://cdsads.u-strasbg.fr/abs/2002A%26A...392..795V} {392, 795}

\bibitem[\protect\citeauthoryear{{Varela} et~al.,}{{Varela}
  et~al.}{2009}]{varela09}
{Varela} J.,  et~al., 2009, \mn@doi [\aap] {10.1051/0004-6361/200809876}, \href
  {http://cdsads.u-strasbg.fr/abs/2009A%26A...497..667V} {497, 667}

\bibitem[\protect\citeauthoryear{{Vikhlinin} et~al.,}{{Vikhlinin}
  et~al.}{2009}]{vikh09III}
{Vikhlinin} A.,  et~al., 2009, \mn@doi [\apj] {10.1088/0004-637X/692/2/1060},
  \href {http://cdsads.u-strasbg.fr/abs/2009ApJ...692.1060V} {692, 1060}

\bibitem[\protect\citeauthoryear{Wall \& Jenkins}{Wall \&
  Jenkins}{2003}]{wall2003}
Wall J.,  Jenkins C.,  2003, Practical Statistics for Astronomers.
Cambridge Observing Handbooks for Research Astronomers, Cambridge University
  Press, \url {https://books.google.it/books?id=ekyupqnDFzMC}

\bibitem[\protect\citeauthoryear{{Wen}, {Han}  \& {Liu}}{{Wen}
  et~al.}{2012}]{wen12}
{Wen} Z.~L.,  {Han} J.~L.,   {Liu} F.~S.,  2012, VizieR Online Data Catalog,
  \href {http://cdsads.u-strasbg.fr/abs/2012yCat..21990034W} {219, 90034}

\bibitem[\protect\citeauthoryear{{Williamson} et~al.,}{{Williamson}
  et~al.}{2011}]{williamson11}
{Williamson} R.,  et~al., 2011, \mn@doi [\apj] {10.1088/0004-637X/738/2/139},
  \href {http://cdsads.u-strasbg.fr/abs/2011ApJ...738..139W} {738, 139}

\bibitem[\protect\citeauthoryear{{Zhang}, {Andernach}, {Caretta}, {Reiprich},
  {B{\"o}hringer}, {Puchwein}, {Sijacki}  \& {Girardi}}{{Zhang}
  et~al.}{2011}]{zhang11}
{Zhang} Y.-Y.,  {Andernach} H.,  {Caretta} C.~A.,  {Reiprich} T.~H.,
  {B{\"o}hringer} H.,  {Puchwein} E.,  {Sijacki} D.,   {Girardi} M.,  2011,
  \mn@doi [\aap] {10.1051/0004-6361/201015830}, \href
  {http://adsabs.harvard.edu/abs/2011A%26A...526A.105Z} {526, A105}

\bibitem[\protect\citeauthoryear{{van Weeren} et~al.,}{{van Weeren}
  et~al.}{2013}]{vanWeeren13}
{van Weeren} R.~J.,  et~al., 2013, \mn@doi [\apj]
  {10.1088/0004-637X/769/2/101}, \href
  {http://cdsads.u-strasbg.fr/abs/2013ApJ...769..101V} {769, 101}

\bibitem[\protect\citeauthoryear{{van den Bosch}, {Weinmann}, {Yang}, {Mo},
  {Li}  \& {Jing}}{{van den Bosch} et~al.}{2005}]{vandenBosch.ea:05}
{van den Bosch} F.~C.,  {Weinmann} S.~M.,  {Yang} X.,  {Mo} H.~J.,  {Li} C.,
  {Jing} Y.~P.,  2005, \mn@doi [\mnras] {10.1111/j.1365-2966.2005.09260.x},
  \href {http://adsabs.harvard.edu/abs/2005MNRAS.361.1203V} {361, 1203}

\makeatother
\end{thebibliography}

\begin{appendix}
\section{Notes on individual objects}
\label{app_A}
In this appendix, we provide some details on the BCG association for a few clusters, where the association may be ambiguous. 
\begin{description}
\item[\textbf{PSZ1 INDEX 26.}] This \Planck\ cluster is associated to A2744, which has 5 bright elliptical galaxies, potentially candidate BCGs \citep{mann_ebe}. We chose the brightest one in the $r$ band in the catalogue of \citet{owers11}.
\item[\textbf{PSZ1 INDEX 122.}]  This \Planck\ cluster is associated to A2142, which hosts two elliptical galaxies with similar brightness. We chose the one indicated in MaxBCG and in \citet{hoffer12,zhang11}. The magnitude of the galaxy indicated by \citet{crawford99} is slightly smaller in the NED database.  
\item[\textbf{PSZ1 INDEX 141.}]  This \Planck\ cluster is associated to A2069, which hosts a substructure to the NE, visible both in the X-rays and in the SDSS image \citep{owers09}. The galaxy associated to the NE subcluster (SDSS J$152424.07+300021.8$) is indicated as BCG in the MaxBCG catalogue. As it is slightly brighter ($14.9 \pm 0.003$ in the $r-$band) than the elliptical galaxy (SDSS J$152408.43+295255.5$) close to the centre of the main cluster ($15.830\pm 0.003$) chosen by \citet{hoffer12}, we keep the MaxBCG selection. 
\item[\textbf{PSZ1 INDEX 185.}] This \Planck\ cluster is associated to A2249. We chose the BCG in the RedMaPPer catalogue \citep{rykoff14}, which coincides with the one chosen by \citet{crawford99} and is brighter than the one indicated by \citet{wen12}.
\item[\textbf{PSZ1 INDEX 187.}]  This \Planck\ cluster is associated to the Coma Cluster, which is well known for having two BCGs. We chose NGC 4874. 
\item[\textbf{PSZ1 INDEX 319.}]  This \Planck\ cluster is associated to A1763. The same BCG is indicated in MaXBCG, RedMaPPer, Wen12 and \citet{hoffer12}. However, we find a separation of 160 kpc, while \citet{mann_ebe} find only one kpc. This is probably due to a different choice of the X-ray peak rather than of the BCG, which is made difficult by the presence of a point source in the central region of this cluster.  
\item[\textbf{PSZ1 INDEX 389.}]  This \Planck\ cluster is associated to the northern component of A1758, a well known double cluster. The BCG chosen by MaxBCG coincides with the choice of RedMaPPer, \citet{hoffer12} and with the brightest galaxy in the catalogue of \citet{boschin12}. The offset is only $1.7$ kpc, although it is a very disturbed object.
\item[\textbf{PSZ1 INDEX 407.}] This \Planck\ cluster is associated to A2256, which hosts a few galaxies with similar brightness. We chose the brightest in the WINGS catalogue \citep{varela09}.
\item[\textbf{PSZ1 INDEX 422.}]  This \Planck\ cluster is associated to A1682. MaxBCG and RedMaPPer provide the same BCG (SDSS J$130645.69+463330.7$) while Wen12 finds another galaxy ( 2MASX J$13064997+4633335$) which almost coincides with the X-ray peak.  We chose the latter source as it is slightly brighter in the r-band ($16.362$ vs $16.497$, SDSS r-model). The resulting offset is consistent with the one reported in \citet{mann_ebe}.
\item[\textbf{PSZ1 INDEX 787.}]  This \Planck\ cluster is associated to A1367. Following \citet{sun09}, we chose NGC3862 as the cluster BCG. 
\item[\textbf{PSZ1 INDEX 988.}]  This \Planck\ cluster is associated to A1553. MaxBCG, Wen12 and \citet{stott12} indicate as BCG SDSS J$123048.87+103246.9$ which is slightly brighter in the $r$ band than the galaxy indicated by RedMaPPer and \citet{crawford99}.
\item[\textbf{PSZ1 INDEX 1182.}]  This \Planck\ cluster is associated to A1835, a well known relaxed cool core cluster, with a clear BCG at its center with a high star formation rate \citep{mcnamara06}. This galaxy is correctly indicated as BCG by \citet{wen12} and \citet{hoffer12}, while MaxBCG and RedMaPPer indicate two other galaxies, possibly because this peculiar star-forming BCG is not classified as a cluster member by their Red Sequence analysis.  

\end{description}	

\section{Impact of projection effects}
\label{app_B}
In this appendix we estimate the fraction of disturbed objects which are classified as relaxed because of projection effects.
Let us consider a cluster where the physical distance between the X-ray peak and the BCG is $r$, larger than the reference projected distance $b_T=0.02R_{500}$ that we used to separate clusters into the two classes. If the line connecting the BCG to the X-ray peak forms a small angle with the line of sight, it is possible that the projected component of the offset in the plane of the sky is smaller than our threshold and therefore be classified as a relaxed cluster. For each physical separation $r>b_T$ we can thus estimate the probability that the cluster can be misclassified as relaxed, with
\begin{equation}
\label{eq:prob}
P(r)=\frac{1}{\pi}\arcsin\left(\frac{b_T}{r}\right).
\end{equation} 
In order to estimate the number of disturbed objects misclassified as relaxed for projection effects we need to convolve the probability function in Eq. \ref{eq:prob} with the 3-D density distribution of offsets, that we derived by deprojecting the observed 2D distribution. 
The distribution of  the fraction of disturbed objects ($N/N_{tot}$, where $N_{tot}=132$, i.e the total number of objects in the \Planck\ sample) with measured projected offsets in each bin as a function of  the projected distance $b$ can be well fitted with a power-law and the best-fit parameters can be used to estimate the parameters of the 3-D distribution which is also a power-law.
In fact, we can assume
\begin{equation}
\label{eq:defpl}
\rho_{offset}(r)=\frac{N(r)}{N_{tot}*V_{shell}}=\rho_\star\left(\frac{r}{r_\star}\right)^\alpha,
\end{equation}
where $\rho_\star$ is the normalization of the power-law at a convenient radius $r_*$ that we chose $=b_T$ and $V_{shell}$ is the volume of a thin sperical shell of radius $r$. We can project this function along the line of sight to be compared with our measured projected density function as:
\begin{equation}
\label{eq:firstint}
\sigma(b)=\frac{N(r)}{N_{tot}*A_{shell}}=\int_{-\infty}^{\infty}\rho(r)dz,
\end{equation}
where $A_{shell}$ is the area of thin shell corresponding to a projected offset $b$ and $z$ is the coordinate representing the line of sight. 
We then substitute Eq. \ref{eq:defpl} into Eq. \ref{eq:firstint} and express $dz$ as a function of $r$ and $b$, and we get
\begin{equation}
\sigma(b)=2\rho_\star\int_{b}^{+\infty}\left(\frac{r}{r_\star}\right)^\alpha\frac{r}{\sqrt{r^2-b^2}}dr .
\end{equation}
Changing variables, we obtain
\begin{equation}
\sigma(b)=\rho_\star r_\star \int_{(b/r_\star)^2}^{+\infty} x^{\alpha/2}\left(x - \frac{b^2}{r_\star^2}\right)^{-1/2}dx,
\end{equation}
which can be integrated using Eq. 3.191.2 in \citet{gradshteyn2007} and we find
\begin{align}
\notag
\sigma(b)=\rho_\star r_\star \left(\frac{b}{r_\star}\right)^{\alpha+1}\beta\left(-\frac{\alpha}{2}-\frac{1}{2},\frac{1}{2}\right)= \\
=\rho_\star r_\star \sqrt{\pi}\frac{\Gamma\left(-\frac{\alpha+1}{2}\right)}{\Gamma\left(-\frac{\alpha}{2}\right)}\left(\frac{b}{r_\star}\right)^{\alpha+1}.
\end{align}
Therefore if we fit our projected distribution with a power-law in the form $\sigma(b)=B_\star*(b/b_T)^\beta$, we can derive from the best-fit parameters ($\beta=\alpha+1=-2.39\pm0.11$) the shape and normalization of the parent 3D distribution $\rho(r)$.
We can then calculate the fraction of disturbed objects which are classified as relaxed by integrating the product $\rho(r)P(r)$ over the volume spanned by the offset, i.e.
\begin{align}
\notag
f=\int_{b_T}^{R_{500}} \rho(r)P(r)4\pi r^2 dr = \\
 =\frac{4\rho_\star}{r_\star^\alpha}\int_{b_T}^{R_{500}}r^{\alpha+2}\arcsin{\left(\frac{b_T}{r}\right)} dr.
\label{eq:final}
\end{align}
We can calculate numerically the definite integral in Eq. \ref{eq:final} and find that the fraction of objects which are classified as relaxed although they are disturbed is $7.5\%$.
The fraction is rather small because the apparent area within which the BCG and the X-ray peak should fall to be classified as relaxed (a circle of radius $0.02R_{500}$) is small if compared to the whole cluster size. We can thus correct the fraction of relaxed objects in the \Planck\ sample to be $45\%$.
\end{appendix}
\setcounter{table}{0}
\clearpage	
\onecolumn

\begin{footnotesize}
\landscape
\begin{longtable}{ccccccccccccc}
\hline

  INDEX & NAME & Alt. Name & $z$ & R.A.$_X$ & Dec.$_X$ &  R.A.$_{BCG}$ & Dec.$_{BCG}$ & Reference BCG & $\Theta_{500}$ & \multicolumn{3}{c}{$D_{X,BCG}$} \\
   &  &  &  &  &  &  &  &  &   (arcmin) & arcsec & kpc & $0.01 R_{500}$\\
\hline
\endhead

\hline
\caption{Properties of the clusters in our sample. Col. [1] is the INDEX in the PSZ1 catalogue, col. [2] the \Planck\ name, col. [3] provides an alternative name and col. [4] the redshift of the cluster. Cols. [5] and [6] are the coordinates of the X-ray peak, while Cols. [7] and [8] are the coordinates of the BCG and Col. [9] the reference we used to associate a BCG to each cluster. Col. [10] is the angular scale corresponding to $R_{500}$ and Cols. [11]-[13] provide our indicator $D_{X-BCG}$ in units of arcsec, kpc and $0.01R_{500}$.}
\endfoot

\caption{Properties of the clusters in our sample. Col. [1] is the INDEX in the PSZ1 catalogue, col. [2] the \Planck\ name, col. [3] provides an alternative name and col. [4] the redshift of the cluster. Cols. [5] and [6] are the coordinates of the X-ray peak, while Cols. [7] and [8] are the coordinates of the BCG and Col. [9] the reference we used to associate a BCG to each cluster. Col. [10] is the angular scale corresponding to $R_{500}$ and Cols. [11]-[13] provide our indicator $D_{X-BCG}$ in units of arcsec, kpc and $0.01R_{500}$.}
\endlastfoot

10 & PSZ1 G003.93-59.42 & RXC J2234.5-3744 & 0.151 & 338.6166 & -37.7297  & 338.61 & -37.744 & \citet{coziol09} & 8.02 & 54.70 & 143.8 & 11.37\\
  17 & PSZ1 G006.45+50.56 & RXC J1510.9+0543 & 0.0766 & 227.7339 & 5.7446  & 227.734 & 5.745 & \citet{coziol09} & 15.00 & 1.56 & 2.3 & 0.17\\
  18 & PSZ1 G006.68-35.52 & RXC J2034.7-3548 & 0.0894 & 308.6865 & -35.8162  & 308.689 & -35.824 & \citet{coziol09} & 10.93 & 29.02 & 48.4 & 4.42\\
  23 & PSZ1 G008.33-64.74 & ACO S 1077 & 0.312 & 344.7013 & -34.8023  & 344.7016 & -34.8022 & \citet{stanford02} & 4.58 & 0.86 & 3.9 & 0.31\\
  24 & PSZ1 G008.42-56.34 & RXC J2217.7-3543 & 0.1486 & 334.4407 & -35.7243  & 334.441 & -35.725 & \citet{coziol09} & 7.32 & 2.78 & 7.2 & 0.63\\
  26 & PSZ1 G009.02-81.22 & RXC J0014.3-3023 & 0.3066 & 3.5814 & -30.3917  & 3.5864 & -30.391 & \citet{owers11} & 4.97 & 15.90 & 71.9 & 5.33\\
  54 & PSZ1 G021.10+33.24 & RXC J1632.7+0534 & 0.1514 & 248.1955 & 5.5758  & 248.1958 & 5.5757 & \citet{zhang11} & 8.49 & 1.08 & 2.8 & 0.21\\
  76 & PSZ1 G029.10+44.54 & RXC J1602.3+1601 & 0.0353 & 240.5709 & 15.9745  & 240.571 & 15.9747 & \citet{hoffer12} & 23.73 & 0.45 & 0.3 & 0.03\\
  92 & PSZ1 G033.43-48.44 & RXC J2152.4-1933 & 0.0943 & 328.0882 & -19.5478  & 328.0915 & -19.5468 & \citet{hoffer12} & 10.45 & 11.46 & 20.1 & 1.83\\
  93 & PSZ1 G033.84+77.17 & RXC J1348.8+2635 & 0.0622 & 207.22 & 26.5899  & 207.219 & 26.593 & \citet{coziol09} & 15.94 & 11.71 & 14.0 & 1.22\\
  94 & PSZ1 G034.03-76.59 & RXC J2351.6-2605 & 0.2264 & 357.9141 & -26.0841  & 357.9142 & -26.0843 & \citet{coziol09} & 5.70 & 0.55 & 2.0 & 0.16\\
  108 & PSZ1 G039.81-39.96 & RXC J2127.1-1209 & 0.176 & 321.788 & -12.1675  & 321.807 & -12.163 & \citet{coziol09} & 6.66 & 68.69 & 204.7 & 17.19\\
  120 & PSZ1 G042.85+56.63 & RXC J1522.4+2742 & 0.0723 & 230.623 & 27.7064  & 230.6216 & 27.7076 & \citet{zhang11} & 13.59 & 6.26 & 8.6 & 0.77\\
  122 & PSZ1 G044.24+48.66 & RXC J1558.3+2713 & 0.0894 & 239.587 & 27.2289  & 239.583 & 27.233 & \citet{maxbcg} & 14.15 & 19.41 & 32.4 & 2.29\\
  140 & PSZ1 G046.48-49.42 & RXC J2210.3-1210 & 0.0846 & 332.5783 & -12.1704  & 332.578 & -12.171 & \citet{coziol09} & 11.89 & 2.49 & 3.9 & 0.35\\
  141 & PSZ1 G046.90+56.48 & RXC J1524.1+2955 & 0.1145 & 231.0345 & 29.8826  & 231.1 & 30.006 & \citet{maxbcg} & 9.61 & 489.13 & 1015.8 & 84.84\\
  153 & PSZ1 G049.22+30.84 & RXC J1720.1+2637 & 0.1644 & 260.0414 & 26.6248  & 260.0418 & 26.6256 & \citet{hoffer12} & 7.32 & 3.21 & 9.1 & 0.73\\
  166 & PSZ1 G053.52+59.52 & RXC J1510.1+3330 & 0.113 & 227.5521 & 33.511  & 227.549 & 33.486 & \citet{maxbcg} & 9.40 & 90.44 & 185.7 & 16.03\\
  174 & PSZ1 G055.58+31.87 & RXC J1722.4+3208 & 0.224 & 260.6136 & 32.1329  & 260.613 & 32.133 & \citet{wen12} & 5.90 & 1.71 & 6.2 & 0.48\\
  177 & PSZ1 G055.95-34.87 & RXC J2135.2+0125 & 0.231 & 323.832 & 1.4248  & 323.828 & 1.424 & \citet{wen12} & 5.63 & 14.64 & 54.0 & 4.33\\
  180 & PSZ1 G056.79+36.30 & RXC J1702.7+3403 & 0.0953 & 255.6774 & 34.0604  & 255.677 & 34.06 & \citet{maxbcg} & 10.28 & 1.82 & 3.2 & 0.29\\
  181 & PSZ1 G056.94-55.06 & RXC J2243.3-0935 & 0.447 & 340.8391 & -9.595  & 340.832 & -9.592 & \citet{rykoff14} & 3.78 & 27.39 & 157.1 & 12.08\\
  183 & PSZ1 G057.28-45.37 & RXC J2211.7-0349 & 0.397 & 332.9411 & -3.827  & 332.941 & -3.829 & \citet{wen12} & 4.01 & 7.16 & 38.3 & 2.98\\
  185 & PSZ1 G057.63+34.92 & RXC J1709.8+3426 & 0.0802 & 257.441 & 34.4551  & 257.4524 & 34.4579 & \citet{rykoff14} & 11.63 & 35.07 & 53.1 & 5.03\\
  187 & PSZ1 G057.84+87.98 & RXC J1259.7+2756 & 0.0231 & 194.9142 & 27.9536  & 194.8988 & 27.9593 & DSS+NED+2MASS & 43.65 & 53.26 & 24.8 & 2.03\\
  207 & PSZ1 G062.94+43.69 & RXC J1628.6+3932 & 0.0299 & 247.1594 & 39.5512  & 247.159 & 39.551 & \citet{coziol09} & 27.39 & 1.23 & 0.7 & 0.08\\
  224 & PSZ1 G067.19+67.44 & RXC J1426.0+3749 & 0.1712 & 216.5134 & 37.8241  & 216.486 & 37.816 & \citet{maxbcg} & 7.29 & 83.09 & 242.2 & 19.00\\
  235 & PSZ1 G071.21+28.86 & RXC J1752.0+4440 & 0.366 & 267.9948 & 44.6636  & 267.9725 & 44.6539 & \citet{wen12} & 3.89 & 67.15 & 341.3 & 28.78\\
  242 & PSZ1 G072.61+41.47 & RXC J1640.3+4642 & 0.228 & 250.0838 & 46.7119  & 250.083 & 46.711 & \citet{maxbcg} & 6.64 & 3.79 & 13.8 & 0.95\\
  248 & PSZ1 G073.98-27.83 & RXC J2153.5+1741 & 0.2329 & 328.4034 & 17.6957  & 328.403 & 17.696 & \citet{hoffer12} & 6.21 & 1.93 & 7.2 & 0.52\\
  252 & PSZ1 G075.71+13.51 & RXC J1921.1+4357 & 0.0557 & 290.3021 & 43.9501  & 290.2921 & 43.9456 & \citet{coziol09} & 21.89 & 30.65 & 33.1 & 2.33\\
  256 & PSZ1 G077.89-26.62 & RXC J2200.8+2058 & 0.147 & 330.219 & 20.9685  & 330.219 & 20.969 & \citet{wen12} & 7.68 & 1.85 & 4.8 & 0.40\\
  268 & PSZ1 G081.01-50.92 & RXC J2311.5+0338 & 0.2998 & 347.8888 & 3.6358  & 347.888 & 3.634 & \citet{wen12} & 4.67 & 7.10 & 31.6 & 2.53\\
  319 & PSZ1 G092.67+73.44 & RXC J1335.3+4059 & 0.2279 & 203.8179 & 41.0  & 203.834 & 41.001 & \citet{maxbcg} & 6.05 & 44.00 & 160.5 & 12.13\\
  325 & PSZ1 G093.93+34.92 & RXC J1712.7+6403 & 0.0809 & 258.1703 & 64.0658  & 258.12 & 64.061 & \citet{wen12} & 13.01 & 81.07 & 123.7 & 10.38\\
  326 & PSZ1 G094.00+27.41 & H1821+643 Cluster & 0.3315 & 275.4888 & 64.3434  & 275.5147 & 64.3836 & \citet{wen12} & 4.07 & 150.40 & 716.9 & 61.62\\
  341 & PSZ1 G097.72+38.13 & RXC J1635.8+6612 & 0.1709 & 248.9618 & 66.2118  & 248.954 & 66.2125 & \citet{crawford99} & 7.10 & 11.64 & 33.9 & 2.73\\
  388 & PSZ1 G106.84-83.24 & RXC J0043.4-2037 & 0.2924 & 10.852 & -20.6239  & 10.8543 & -20.6182 & \citet{hoffer12} & 5.09 & 22.00 & 96.2 & 7.20\\
  389 & PSZ1 G107.14+65.29 & RXC J1332.7+5032 & 0.2799 & 203.1598 & 50.56  & 203.16 & 50.56 & \citet{maxbcg} & 5.04 & 0.39 & 1.7 & 0.13\\
  407 & PSZ1 G110.99+31.74 & RXC J1703.8+7838 & 0.0581 & 255.8017 & 78.6496  & 256.1121 & 78.6406 & \citet{varela09} & 19.01 & 222.38 & 250.2 & 19.50\\
  411 & PSZ1 G112.48+57.02 & RXC J1336.1+5912 & 0.0701 & 204.0312 & 59.2031  & 204.035 & 59.206 & \citet{coziol09} & 12.65 & 12.48 & 16.7 & 1.64\\
  415 & PSZ1 G113.84+44.33 & RXC J1414.2+7115 & 0.225 & 213.4809 & 71.2987  & 213.4879 & 71.296 & DSS+NED+2MASS & 5.17 & 12.61 & 45.6 & 4.07\\
  422 & PSZ1 G114.99+70.36 & RXC J1306.9+4633 & 0.2259 & 196.7084 & 46.5588  & 196.7083 & 46.5593 & \citet{wen12} & 5.53 & 1.77 & 6.4 & 0.53\\
  423 & PSZ1 G115.20-72.07 & RXC J0041.8-0918 & 0.0555 & 10.46 & -9.303  & 10.461 & -9.303 & \citet{wen12} & 18.22 & 3.46 & 3.7 & 0.32\\
  454 & PSZ1 G124.20-36.47 & RXC J0055.9+2622 & 0.1971 & 13.96 & 26.4101  & 13.9604 & 26.4108 & \citet{crawford99} & 6.53 & 2.83 & 9.2 & 0.72\\
  459 & PSZ1 G125.68-64.12 & RXC J0056.3-0112 & 0.0442 & 14.0663 & -1.2553  & 14.0671 & -1.2554 & \citet{varela09} & 19.93 & 3.12 & 2.7 & 0.26\\
  502 & PSZ1 G139.17+56.37 & RXC J1142.5+5832 & 0.322 & 175.6004 & 58.5329  & 175.603 & 58.535 & \citet{wen12} & 4.34 & 9.07 & 42.4 & 3.48\\
  513 & PSZ1 G143.28+65.22 & RXC J1159.2+4947 & 0.3633 & 179.8089 & 49.7944  & 179.812 & 49.797 & \citet{wen12} & 4.04 & 11.65 & 58.9 & 3.43\\
  530 & PSZ1 G149.21+54.17 & RXC J1058.4+5647 & 0.1369 & 164.5966 & 56.7948  & 164.599 & 56.795 & \citet{wen12} & 8.55 & 4.80 & 11.6 & 0.94\\
  532 & PSZ1 G149.55-84.16 & RXC J0102.7-2152 & 0.0569 & 15.674 & -21.8805  & 15.674 & -21.8823 & \citet{coziol09} & 15.25 & 6.41 & 7.1 & 0.70\\
  533 & PSZ1 G149.75+34.68 & RXC J0830.9+6551 & 0.1818 & 127.7468 & 65.8389  & 127.7388 & 65.8422 & \citet{crawford99} & 7.31 & 16.80 & 51.4 & 3.83\\
  558 & PSZ1 G159.81-73.47 & RXC J0131.8-1336 & 0.206 & 22.969 & -13.6108  & 22.9688 & -13.6114 & \citet{stott08} & 6.55 & 2.35 & 7.9 & 0.60\\
  567 & PSZ1 G163.69+53.52 & RXC J1022.5+5006 & 0.158 & 155.6177 & 50.1066  & 155.618 & 50.106 & \citet{wen12} & 6.97 & 2.30 & 6.3 & 0.55\\
  572 & PSZ1 G165.06+54.13 & RXC J1023.6+4907 & 0.144 & 155.9152 & 49.1447  & 155.916 & 49.144 & \citet{wen12} & 7.41 & 3.32 & 8.4 & 0.75\\
  578 & PSZ1 G166.11+43.40 & RXC J0917.8+5143 & 0.2172 & 139.4631 & 51.7255  & 139.4726 & 51.7271 & \citet{rykoff14} & 5.97 & 22.02 & 77.5 & 6.14\\
  582 & PSZ1 G167.64+17.63 & RXC J0638.1+4747 & 0.174 & 99.5151 & 47.7986  & 99.5165 & 47.7982 & \citet{crawford95} & 7.03 & 3.54 & 10.5 & 0.84\\
  608 & PSZ1 G180.25+21.03 & RXC J0717.5+3745 & 0.546 & 109.3822 & 37.7577  & 109.398  & 48.697 & \citet{mann_ebe} & 3.38 & 62.77 & 401.0 & 30.91\\ 
  628 & PSZ1 G186.37+37.26 & RXC J0842.9+3621 & 0.282 & 130.7398 & 36.366  & 130.74 & 36.366 & \citet{maxbcg} & 5.65 & 0.62 & 2.6 & 0.18\\
  655 & PSZ1 G195.78-24.29 & RXC J0454.1+0255 & 0.203 & 73.5296 & 2.9065  & 73.516 & 2.8925 & \citet{crawford99} & 6.32 & 70.30 & 234.7 & 18.54\\
  681 & PSZ1 G205.94-39.46 & RXC J0417.5-1154 & 0.443 & 64.3945 & -11.9091  &  64.394 & -11.909  & \citet{mann_ebe} & 4.00 & 1.11 & 6.3 & 0.46\\ 
  715 & PSZ1 G216.60+47.00 & RXC J0949.8+1707 & 0.3826 & 147.4661 & 17.1194  & 147.465 & 17.120  & \citet{mann_ebe} & 3.97 & 1.39 & 7.3 & 0.58\\ 
  757 & PSZ1 G225.91-19.98 & RXC J0600.1-2007 & 0.46 & 90.0335 & -20.1362  & 90.034 & -20.1358 & \citet{story11} & 3.73 & 2.10 & 12.2 & 0.94\\
  758 & PSZ1 G226.19+76.78 & RXC J1155.3+2324 & 0.1427 & 178.8246 & 23.4061  & 178.825 & 23.405 & \citet{maxbcg} & 8.13 & 4.11 & 10.3 & 0.84\\
  759 & PSZ1 G226.19-21.92 & RXC J0552.8-2103 & 0.0989 & 88.2133 & -21.0527  & 88.213 & -21.051 & \citet{coziol09} & 10.09 & 6.22 & 11.4 & 1.03\\
  772 & PSZ1 G229.23-17.23 & RXC J0616.3-2156 & 0.171 & 94.1031 & -21.9377  & 94.101 & -21.935 & DSS+2MASS+NED & 6.92 & 11.77 & 34.3 & 2.84\\
  773 & PSZ1 G229.70+77.97 & RXC J1201.3+2306 & 0.269 & 180.3062 & 23.1115  & 180.32 & 23.109 & \citet{maxbcg} & 5.15 & 46.70 & 192.6 & 15.11\\
  774 & PSZ1 G229.92+15.28 & RXC J0817.4-0730 & 0.0704 & 124.3551 & -7.5111  & 124.357 & -7.512 & \citet{wen12} & 14.35 & 7.47 & 10.0 & 0.87\\
  787 & PSZ1 G234.54+73.03 & RXC J1144.6+1945 & 0.0214 & 176.1723 & 19.711  & 176.2709 & 19.6063 & \citet{sun09} & 32.59 & 503.65 & 218.0 & 25.76\\
  796 & PSZ1 G236.93-26.65 & RXC J0547.6-3152 & 0.1483 & 86.9071 & -31.8732  & 86.9073 & -31.8734 & \citet{hoffer12} & 7.55 & 0.89 & 2.3 & 0.20\\
  801 & PSZ1 G239.29+24.75 & RXC J0909.1-0939 & 0.0542 & 137.3199 & -9.6885  & 137.135 & -9.63 & \citet{coziol09} & 20.67 & 689.18 & 726.4 & 55.58\\
  802 & PSZ1 G239.30-26.01 & RXC J0553.4-3342 & 0.43 & 88.3669 & -33.7091  & 88.357 & -33.708 & \citet{mann_ebe} & 3.80 & 29.15 & 163.5 & 12.8\\ 
  815 & PSZ1 G241.75-30.89 & RXC J0532.9-3701 & 0.2708 & 83.2329 & -37.0264  & 83.2167 & -37.0341 & DSS+2MASS+NED & 4.90 & 54.36 & 225.3 & 18.48\\
  816 & PSZ1 G241.76-24.01 & RXC J0605.8-3518 & 0.1392 & 91.4743 & -35.302  & 91.475 & -35.302 & \citet{coziol09} & 8.04 & 2.13 & 5.2 & 0.44\\
  818 & PSZ1 G241.98+14.87 & RXC J0841.9-1729 & 0.1687 & 130.4663 & -17.4628  & 130.47 & -17.468 & \citet{vanWeeren13} & 7.21 & 22.72 & 65.4 & 5.25\\
  824 & PSZ1 G243.60+67.74 & RXC J1132.8+1428 & 0.0834 & 173.2127 & 14.4568  & 173.213 & 14.461 & \citet{wen12} & 11.75 & 15.15 & 23.7 & 2.15\\
  826 & PSZ1 G244.35-32.15 & RXC J0528.9-3927 & 0.2839 & 82.221 & -39.471  & 82.221 & -39.472 & \citet{hoffer12} & 4.84 & 3.67 & 15.7 & 1.26\\
  838 & PSZ1 G246.53-26.07 & RXC J0601.7-3959 & 0.0468 & 90.549 & -39.9497  & 90.172 & -40.044 & \citet{coziol09} & 16.59 & 1093.65 & 1004.0 & 109.87\\
  857 & PSZ1 G250.92-36.24 & RXC J0510.2-4519 & 0.2 & 77.5712 & -45.3208  & 77.573 & -45.322 & \citet{coziol09} & 6.04 & 6.17 & 20.4 & 1.70\\
  862 & PSZ1 G252.99-56.06 & RXC J0317.9-4414 & 0.0752 & 49.4903 & -44.2381  & 49.49 & -44.238 & \citet{coziol09} & 11.71 & 0.94 & 1.3 & 0.13\\
  877 & PSZ1 G255.60-46.18 & SPT-CLJ0411-4819 & 0.4235 & 62.8186 & -48.3154  & 62.7957 & -48.3276 & \citet{story11} & 3.37 & 70.20 & 390.3 & 31.02\\
  880 & PSZ1 G256.55-65.69 & RXC J0225.9-4154 & 0.2195 & 36.4717 & -41.9162  & 36.471 & -41.914 & DSS+NED+2MASS & 5.54 & 8.25 & 29.3 & 2.48\\
  882 & PSZ1 G257.32-22.19 & RXC J0637.3-4828 & 0.2026 & 99.3108 & -48.4717  & 99.311 & -48.473 & DSS+2MASS+NED & 5.57 & 4.88 & 16.3 & 1.46\\
  889 & PSZ1 G260.00-63.45 & RXC J0232.2-4420 & 0.2836 & 38.0782 & -44.3463  & 38.083 & -44.351 & \citet{hoffer12} & 4.73 & 20.93 & 89.6 & 7.37\\
  898 & PSZ1 G262.27-35.38 & RXC J0516.6-5430 & 0.2952 & 79.1531 & -54.5125  & 79.156 & -54.5 & \citet{coziol09} & 5.03 & 45.43 & 200.1 & 15.07\\
  901 & PSZ1 G262.72-40.92 &  & 0.421 & 69.5718 & -54.3233  & 69.573 & -54.322 & \citet{menanteau10} & 3.60 & 5.27 & 29.2 & 2.44\\
  904 & PSZ1 G263.14-23.42 & RXC J0638.7-5358 & 0.2266 & 99.7025 & -53.974  & 99.688 & -53.973 & \citet{rawle12} & 5.66 & 30.99 & 112.6 & 9.12\\
  905 & PSZ1 G263.19-25.22 & RXC J0627.2-5428 & 0.0506 & 96.698 & -54.5465  & 96.8597 & -54.5173 & \citet{postman95} & 16.28 & 353.78 & 349.6 & 36.22\\
  907 & PSZ1 G263.68-22.55 & RXC J0645.4-5413 & 0.1644 & 101.3707 & -54.2288  & 101.373 & -54.227 & \citet{coziol09} & 7.87 & 8.04 & 22.7 & 1.70\\
  914 & PSZ1 G265.02-48.96 & RXC J0342.8-5338 & 0.059 & 55.7135 & -53.6299  & 55.721 & -53.631 & \citet{coziol09} & 16.33 & 16.48 & 18.8 & 1.68\\
  920 & PSZ1 G266.02-21.23 & RXC J0658.5-5556 & 0.2965 & 104.5847 & -55.9419  & 104.6468 & -55.9492 & \citet{hoffer12} & 5.57 & 127.99 & 565.5 & 38.29\\
  924 & PSZ1 G266.85+25.06 & RXC J1023.8-2715 & 0.2542 & 155.9592 & -27.2566  & 155.96 & -27.256 & \citet{hoffer12} & 5.37 & 3.48 & 13.8 & 1.08\\
  939 & PSZ1 G271.18-30.95 &  & 0.37 & 87.3341 & -62.0876  & 87.333 & -62.087 & \citet{williamson11} & 3.90 & 2.92 & 14.9 & 1.25\\
  944 & PSZ1 G272.08-40.16 & RXC J0431.4-6126 & 0.0589 & 67.8054 & -61.4535  & 67.806 & -61.453 & \citet{song12} & 19.11 & 2.04 & 2.3 & 0.18\\
  951 & PSZ1 G273.54+63.23 & RXC J1200.4+0320 & 0.1339 & 180.1064 & 3.3471  & 180.106 & 3.347 & \citet{maxbcg} & 8.46 & 1.48 & 3.5 & 0.29\\
  958 & PSZ1 G277.75-51.71 &  & 0.438 & 43.569 & -58.9488  & 43.537 & -58.972 & \citet{song12} & 3.60 & 102.55 & 581.5 & 47.47\\
  960 & PSZ1 G278.58+39.15 & RXC J1131.9-1955 & 0.3075 & 172.9774 & -19.929  & 172.976 & -19.9279 & \citet{hoffer12} & 4.83 & 6.33 & 28.7 & 2.19\\
  971 & PSZ1 G280.21+47.83 & RXC J1149.7-1219 & 0.1557 & 177.4419 & -12.3164  & 177.441 & -12.314 & \citet{coziol09} & 7.28 & 9.19 & 24.8 & 2.11\\
  980 & PSZ1 G282.45+65.18 & RXC J1217.6+0339 & 0.0766 & 184.4238 & 3.6544  & 184.421 & 3.656 & \citet{wen12} & 13.24 & 11.55 & 16.8 & 1.46\\
  984 & PSZ1 G284.43+52.44 & RXC J1206.2-0848 & 0.4414 & 181.5511 & -8.8007  & 181.551 & -8.801 & \citet{ebeling09} & 3.84 & 1.31 & 7.5 & 0.57\\
  986 & PSZ1 G285.50-62.25 & RXC J0145.0-5300 & 0.1168 & 26.2429 & -53.0226  & 26.4563 & -52.9965 & DSS+2MASS+NED & 8.14 & 471.53 & 996.3 & 96.57\\
  988 & PSZ1 G285.63+72.72 & RXC J1230.7+1033 & 0.165 & 187.6978 & 10.5528  & 187.7036 & 10.5464 & \citet{maxbcg} & 7.04 & 31.07 & 87.9 & 7.36\\
  994 & PSZ1 G286.60-31.23 &  & 0.21 & 82.8698 & -75.1795  & 82.876 & -75.184 & DSS+2MASS+NED & 5.57 & 17.15 & 58.8 & 5.13\\
  998 & PSZ1 G287.00+32.90 &  & 0.39 & 177.7059 & -28.0736  & 177.709 & -28.082 & \citet{bonafede14} & 4.66 & 31.93 & 168.9 & 11.42\\
  1009 & PSZ1 G288.26+39.94 & RXC J1203.2-2131 & 0.1992 & 180.8173 & -21.5337  & 180.8196 & -21.5483 & \citet{valtchanov02} & 6.50 & 53.33 & 175.4 & 13.67\\
  1011 & PSZ1 G288.63-37.67 & RXC J0352.4-7401 & 0.127 & 58.1295 & -74.0336  & 58.123 & -74.031 & \citet{coziol09} & 9.25 & 11.35 & 25.8 & 2.05\\
  1032 & PSZ1 G294.68-37.01 & RXC J0303.7-7752 & 0.2742 & 45.9409 & -77.8794  & 45.9426 & -77.8787 & DSS+2MASS+NED & 4.89 & 2.92 & 12.2 & 1.00\\
  1037 & PSZ1 G295.34+23.34 & RXC J1215.4-3900 & 0.119 & 183.8533 & -39.0363  & 183.864 & -39.033 & DSS+2MASS+NED & 8.63 & 32.26 & 69.3 & 6.23\\
  1041 & PSZ1 G296.42-32.49 & RXC J0351.1-8212 & 0.0613 & 57.8867 & -82.2195  & 57.891 & -82.22 & \citet{coziol09} & 13.27 & 2.76 & 3.3 & 0.35\\
  1046 & PSZ1 G297.94-67.76 & SPT-CLJ0102-49151 & 0.87 & 15.7428 & -49.2742  & 15.7188 & -49.2494 & \citet{menanteau10} & 2.27 & 105.56 & 814.2 & 77.47\\
  1057 & PSZ1 G303.73+33.69 & RXC J1254.6-2913 & 0.0544 & 193.6704 & -29.2272  & 193.671 & -29.227 & \citet{hoffer12} & 16.19 & 2.08 & 2.2 & 0.21\\
  1062 & PSZ1 G304.44+32.45 & RXC J1257.2-3022 & 0.0554 & 194.3423 & -30.364  & 194.341 & -30.363 & \citet{postman95} & 15.65 & 5.31 & 5.7 & 0.57\\
  1065 & PSZ1 G304.86-41.40 &  & 0.41 & 7.0261 & -75.6359  & 7.0379 & -75.6292 & \citet{story11} & 3.67 & 26.36 & 143.7 & 11.98\\
  1066 & PSZ1 G304.91+45.46 & RXC J1257.1-1724 & 0.0473 & 194.2986 & -17.4091  & 194.299 & -17.41 & \citet{coziol09} & 19.56 & 3.42 & 3.2 & 0.29\\
  1079 & PSZ1 G306.77+58.62 & RXC J1259.3-0411 & 0.0845 & 194.8451 & -4.1961  & 194.844 & -4.196 & \citet{coziol09} & 12.51 & 4.05 & 6.4 & 0.54\\
  1095 & PSZ1 G311.98+30.73 & RXC J1327.9-3130 & 0.048 & 201.9869 & -31.4955  & 201.987 & -31.496 & \citet{coziol09} & 20.21 & 1.85 & 1.7 & 0.15\\
  1105 & PSZ1 G313.33+61.13 & RXC J1311.5-0120 & 0.1832 & 197.8734 & -1.3413  & 197.873 & -1.341 & \citet{maxbcg} & 7.45 & 1.76 & 5.4 & 0.39\\
  1109 & PSZ1 G313.88-17.12 & RXC J1601.7-7544 & 0.153 & 240.4511 & -75.7554  & 240.4506 & -75.746 & DSS+2MASS+NED & 8.26 & 33.68 & 89.5 & 6.79\\
  1117 & PSZ1 G315.69-18.05 & RXC J1631.6-7507 & 0.105 & 247.8377 & -75.115  & 247.8713 & -75.1147 & DSS+2MASS+NED & 11.03 & 31.02 & 59.7 & 4.69\\
  1118 & PSZ1 G316.33+28.55 & RXC J1347.4-3250 & 0.0391 & 206.8685 & -32.8646  & 206.868 & -32.865 & \citet{coziol09} & 25.09 & 2.11 & 1.6 & 0.14\\
  1126 & PSZ1 G321.98-47.96 & RXC J2249.9-6425 & 0.094 & 342.491 & -64.4284  & 342.493 & -64.43 & \citet{hoffer12} & 10.67 & 6.64 & 11.6 & 1.04\\
  1134 & PSZ1 G324.05+48.79 & RXC J1347.5-1144 & 0.4516 & 206.8775 & -11.7528  & 206.8777 & -11.7525 & \citet{bildfell08} & 3.81 & 1.33 & 7.6 & 0.58\\
  1136 & PSZ1 G324.51-44.98 & RXC J2218.0-6511 & 0.0951 & 334.5006 & -65.181  & 334.502 & -65.181 & \citet{guzzo09} & 9.88 & 2.09 & 3.7 & 0.35\\
  1157 & PSZ1 G332.21-46.38 & RXC J2201.9-5956 & 0.098 & 330.4717 & -59.9453  & 330.47 & -59.947 & \citet{coziol09} & 11.39 & 6.85 & 12.4 & 1.00\\
  1160 & PSZ1 G332.87-19.26 & RXC J1813.3-6127 & 0.147 & 273.3051 & -61.4511  & 273.3163 & -61.4558 & DSS+2MASS+NED & 7.86 & 25.60 & 65.8 & 5.43\\
  1164 & PSZ1 G335.57-46.47 & RXC J2154.1-5751 & 0.076 & 328.5175 & -57.8674  & 328.518 & -57.868 & \citet{hoffer12} & 12.84 & 2.29 & 3.3 & 0.30\\
  1165 & PSZ1 G336.61-55.43 & RXC J2246.3-5243 & 0.0965 & 341.6111 & -52.7405  & 341.564 & -52.724 & \citet{postman95} & 10.39 & 118.65 & 212.0 & 19.04\\
  1182 & PSZ1 G340.37+60.57 & RXC J1401.0+0252 & 0.2528 & 210.2581 & 2.8787  & 210.2586 & 2.8785 & \citet{wen12} & 5.58 & 1.79 & 7.1 & 0.53\\
  1184 & PSZ1 G340.86-33.36 & RXC J2012.5-5649 & 0.0556 & 303.1373 & -56.8456  & 303.113 & -56.827 & \citet{coziol09} & 19.20 & 82.37 & 88.9 & 7.15\\
  1185 & PSZ1 G340.94+35.10 & RXC J1459.4-1811 & 0.2357 & 224.8712 & -18.1789  & 224.87 & -18.179 & \citet{coziol09} & 5.74 & 4.02 & 15.1 & 1.17\\
  1200 & PSZ1 G346.61+35.06 & RXC J1514.9-1523 & 0.2226 & 228.7583 & -15.3892  & 228.7457 & -15.357 & DSS+2MASS+NED & 6.18 & 124.04 & 444.6 & 33.45\\
  1208 & PSZ1 G349.46-59.92 & RXC J2248.7-4431 & 0.3475 & 342.1846 & -44.5301  & 342.183 & -44.531 & \citet{coziol09} & 4.77 & 5.14 & 25.3 & 1.79\\
  1218 & PSZ1 G356.18-76.06 & RXC J2357.0-3445 & 0.0475 & 359.2544 & -34.7592  & 359.2537 & -34.7558 & \citet{coziol09} & 16.44 & 12.69 & 11.8 & 1.29\\
\hline
\label{tab:maintable}
\end{longtable}
\endlandscape
\end{footnotesize}

\bsp	
\label{lastpage}

\end{document}